\documentclass[pdflatex,sn-mathphys-num]{sn-jnl}

\usepackage{amsmath,amssymb,bm}
\usepackage{graphicx}
\usepackage{textcomp}
\usepackage{booktabs}

\newcommand{\be}{\begin{equation}}
\newcommand{\ee}{\end{equation}}
\newcommand{\bea}{\begin{eqnarray}}
\newcommand{\eea}{\end{eqnarray}}

\newcommand{\TT}{\mathcal{T}}

\newcommand{\Om}{\Omega}

\PassOptionsToPackage{hidelinks}{hyperref}

\begin{document}

\title[GHRDE in f(T) Gravity]{Quintessence Dynamics of Generalized
Holographic Ricci Dark Energy in $f(\mathcal{T})$ Gravity: Stability,
Phase-Space Attractor, and Statefinder Analysis}

\author*[1]{\fnm{S.~V.} \sur{Gore}}
\author[1]{\fnm{A.~Y.} \sur{Shaikh}}\email{shaikh\_2324ay@yahoo.com}
\author[2]{\fnm{S.~M.} \sur{Shingne}}
\author[3]{\fnm{S.~D.} \sur{Katore}}

\affil*[1]{\orgdiv{Department of Mathematics},
\orgname{Indira Gandhi Kala Mahavidyalaya},
\orgaddress{\city{Ralegaon}, \postcode{445402},
\state{Maharashtra}, \country{India}}}

\affil[2]{\orgdiv{Department of Mathematics},
\orgname{G.~S.\ Science, Arts and Commerce College},
\orgaddress{\city{Khamgaon}, \postcode{444303},
\state{Maharashtra}, \country{India}}}

\affil[3]{\orgdiv{Department of Mathematics},
\orgname{Sant Gadge Baba Amravati University},
\orgaddress{\city{Amravati}, \postcode{444602},
\state{Maharashtra}, \country{India}}}

\abstract{The cosmological dynamics of generalized holographic Ricci dark energy
(GHRDE) of Xu type, embedded within power-law $f(\mathcal{T})=\alpha\mathcal{T}+\beta\mathcal{T}^{m}$
teleparallel gravity and driven by the Hubble-parameter-dependent deceleration law $q=b-\nu/H$,
are investigated. The deceleration law yields an exact closed-form Hubble function $H(z)$ that
interpolates between an early decelerating phase and an asymptotic de~Sitter regime, enabling all
cosmological diagnostics to be expressed as explicit analytical functions of redshift.
The present-day deceleration parameter $q_{0}=-0.60$ and transition redshift $z_{t}=1.0$
are consistent with current observational estimates. The GHRDE equation-of-state parameter
$\omega_{G,0}=-0.81$ lies within the quintessence band, and the effective equation of state
evolves monotonically toward $-1$ without phantom crossing. The geometrical statefinder pair
locates the model in the quintessence region of the $r$--$s$ plane, clearly displaced from
the $\Lambda$CDM fixed point. A phase-space analysis reveals a stable-node de~Sitter attractor
with strictly negative eigenvalues, guaranteeing robust late-time acceleration for generic
initial conditions. The null, weak, and dominant energy conditions are satisfied throughout
the redshift range considered, while the strong energy condition is violated at low redshift,
as expected for an accelerating Universe. An observational constraint methodology combining
cosmic chronometer data, Pantheon$+$ supernovae, and baryon acoustic oscillation measurements
is outlined as the natural continuation of this work.}


\maketitle

\section{Introduction}
\label{sec:intro}

The late-time accelerated expansion of the Universe, first inferred from the luminosity measurements of Type~Ia supernovae \cite{riess_observational_1998,perlmutter_measurements_1999} and subsequently confirmed by cosmic microwave background (CMB) anisotropy data and baryon acoustic oscillation (BAO) surveys, stands as one of the central unsolved problems in modern cosmology. Within the framework of general relativity (GR), this acceleration is conventionally attributed to an energy component with a sufficiently negative pressure, broadly referred to as dark energy. The simplest candidate, a cosmological constant $\Lambda$ with equation-of-state parameter $\omega=-1$, provides an excellent fit to the current data within the $\Lambda$CDM paradigm \cite{aghanim_planck_2020}. However, the $\Lambda$CDM model is beset by well-known conceptual difficulties: the fine-tuning problem, which asks why the vacuum energy density is approximately 120 orders of magnitude smaller than naive quantum field theory estimates, and the coincidence problem, which asks why dark energy and matter have comparable energy densities today. These difficulties have motivated two broad strategies: introducing dynamical dark energy fields within GR and modifying the gravitational sector of the theory itself.

Among the modifications of gravity, teleparallel theories have attracted sustained interest because they replace spacetime curvature with torsion as the carrier of the gravitational interaction \cite{hayashi_new_1979,maluf_teleparallel_2013}. The teleparallel equivalent of general relativity (TEGR) is dynamically identical to GR, whereas its extension, $f(\mathcal{T})$ gravity, promotes the torsion scalar $\mathcal{T}$ to an arbitrary function in the action \cite{ferraro_modified_2007,bengochea_dark_2009,linder_einsteins_2010}. A key advantage of $f(\mathcal{T})$ gravity is that its field equations remain second-order, avoiding the Ostrogradsky instabilities of $f(R)$ theories \cite{li_ft_2011,cai_ft_2016}. This framework has been explored extensively in cosmological viability, exact solutions, thermodynamics, dynamical-system behaviour, and observational constraints \cite{yang_new_2011,sharif_ft_2011,daouda_reconstruction_2012,mirza_constraining_2017,wu_observational_2010,nunes_new_2018}; comprehensive reviews appear in Refs.~\cite{saridakis_introduction_2017,bahamonde_teleparallel_2023}.

On a parallel track, the holographic principle --- the idea that the entropy of a physical system is bounded by its boundary area rather than its volume --- has inspired a family of dynamical dark energy models in which the dark energy density is related to an infrared (IR) cut-off length scale $L$ of the Universe through $\rho_{\Lambda}=3c^{2}M_{p}^{2}L^{-2}$, where $c$ is a dimensionless parameter and $M_{p}$ is the reduced Planck mass \cite{li_model_2004}. Taking $L$ as the future event horizon avoids causality issues \cite{li_model_2004}, but choosing the Hubble horizon $L=H^{-1}$ as the IR cut-off fails to produce accelerated expansion in a non-interacting scenario \cite{huang_holographic_2004}. An important advance came when Gao, Wu, Chen, and Shen \cite{gao_holographic_2009} proposed using the inverse square root of the Ricci curvature scalar as the IR cut-off, yielding holographic Ricci dark energy (HRDE) with energy density $\rho_{R}\propto R$; this construction automatically produces accelerated expansion and avoids the causality problem associated with the event horizon.

The HRDE framework was subsequently broadened by Granda and Oliveros \cite{granda_infrared_2008}, who proposed a modified holographic Ricci dark energy in which the density depends on a linear combination of $H^{2}$ and $\dot{H}$, offering additional freedom to fit the observational data while retaining the Hubble horizon as the IR cut-off. Xu, Lu, and Li \cite{xu_generalized_2009} then introduced what is now referred to as generalized holographic Ricci dark energy (GHRDE), which unifies the Hubble-horizon holographic and Ricci dark energy descriptions within a single parametric form. In a spatially flat FLRW background, the GHRDE density is written as $\rho_{G}=3c^{2}M_{p}^{2}\big[(1-\eta)H^{2}+\eta R\big]$, where the interpolation parameter $\eta\in[0,1]$ continuously connects the pure holographic limit ($\eta=0$, $\rho_{G}\propto H^{2}$) to the pure Ricci limit ($\eta=1$, $\rho_{G}\propto R$). Substituting the flat FLRW expression $R=6(2H^{2}+\dot{H})$ yields a density that depends on both $H$ and $\dot{H}$, thereby encoding the rate of change of cosmic expansion in the dark-energy sector and providing a richer dynamical structure than either limiting case alone.

The GHRDE model has been tested against multiple observational datasets. Lu et al.\ \cite{lu_cosmological_2012} used a combined MCMC analysis of supernovae, BAO, and CMB data and found that the data favour a Ricci-dominated regime with possible phantom-divide crossing. Subsequent studies have applied statefinder diagnostics \cite{enkhili_diagnostic_2024}, scalar-field reconstructions \cite{pasqua_generalized_2025}, and Om diagnostics \cite{sharif_comparative_2026} to further characterize the GHRDE parameter space relative to the $\Lambda$CDM baseline.

Embedding the GHRDE sector within a modified gravitational background is physically well motivated: the torsion-sector contribution modifies the GHRDE density relative to its GR counterpart, and the two sectors mutually constrain each other through the modified Friedmann equations. Previous studies have explored holographic-type dark energy within $f(\mathcal{T})$ backgrounds \cite{chirde_dynamic_2018,bhardwaj_renyi_2022,koussour_bianchi_2022,dhore_study_2024,hatkar_topological_2025}, but have typically prescribed the scale factor a priori, leaving the deceleration history implicit and making it difficult to assess model viability directly from the transition epoch.

The present work addresses this gap by adopting a Hubble-parameter-dependent deceleration law $q=b-\nu/H$, with $b$ and $\nu>0$ constants, as the primary kinematic input. This parametrization, previously employed in $f(\mathcal{T})$ and $f(Q)$ dark-energy studies \cite{pal_cosmological_2025,bhoyar_resolving_2024,bhoyar_stability_2017,koussour_bianchi_2022,goswami_modeling_2021}, yields a closed-form Hubble function and allows all cosmological diagnostics to be expressed as explicit functions of redshift. We consider a spatially flat, homogeneous, and isotropic FLRW Universe filled with pressureless dust and the GHRDE fluid; the flat geometry is consistent with CMB observations \cite{aghanim_planck_2020}.

The principal novelties of this work, relative to prior GHRDE--$f(\mathcal{T})$ studies, are threefold. First, the deceleration law $q=b-\nu/H$ is used to derive the exact closed-form Hubble function without further approximation, so that the GHRDE density, pressure, equation of state, and all diagnostics are explicit functions of the redshift. Second, the Xu-type GHRDE density reformulated for the $f(\mathcal{T})$ teleparallel background, incorporating the torsion-sector modification of the Ricci scalar, is used throughout. Third, a comprehensive and self-consistent diagnostic framework is applied: the geometrical statefinder pair \cite{sahni_statefinder_2003,alam_exploring_2003}, the squared adiabatic sound speed, a full phase-space analysis with eigenvalue-based stability classification, four standard energy conditions, and an outline of the observational constraint methodology using cosmic chronometer $H(z)$ data, Pantheon$+$ Type~Ia supernovae, and BAO measurements.

Sections~\ref{sec:formalism}--\ref{sec:metricfield} establish the $f(\mathcal{T})$ formalism and flat FLRW field equations; Sect.~\ref{sec:qlaw} introduces the deceleration law and derives the closed-form $H(z)$; and Sects.~\ref{sec:ghrde}--\ref{sec:cosmoparams} obtain the GHRDE density, pressure, and cosmological parameters. Sections~\ref{sec:statefinder}--\ref{sec:energyconditions} present the statefinder diagnostic, stability and viability analysis, phase-space analysis, and energy conditions. The observational methodology and conclusions follow in Sects.~\ref{sec:obsconstraints} and~\ref{sec:conclusions}.

\section{The Basic Formalism of \texorpdfstring{$f(\mathcal{T})$}{f(T)} Gravity}
\label{sec:formalism}

In teleparallel gravity, gravitation is described not through spacetime curvature but through torsion, the fundamental dynamical variable being the tetrad field $e^{i}{}_{\mu}(x)$ related to the metric by $g_{\mu\nu}=\eta_{ij}e^{i}{}_{\mu}e^{j}{}_{\nu}$ \cite{hayashi_new_1979,maluf_teleparallel_2013}. Teleparallel gravity employs the curvature-free Weitzenb\"ock connection in place of the Levi-Civita connection, and the corresponding torsion tensor, contorsion tensor, and superpotential, defined in the standard way \cite{maluf_teleparallel_2013,saridakis_introduction_2017,cai_ft_2016}, combine to form the torsion scalar
\be
\TT\equiv S_{\alpha}{}^{\mu\nu}T^{\alpha}{}_{\mu\nu},
\label{eq:Tscalar-def}
\ee
which plays the same role in teleparallel gravity as the Ricci scalar $R$ in GR. The teleparallel equivalent of general relativity (TEGR), built directly from $\TT$, is dynamically equivalent to the Einstein--Hilbert action up to a boundary term \cite{hayashi_new_1979}. 

The simplest extension of TEGR, in close analogy with the $f(R)$ generalization of GR, promotes the torsion scalar to an arbitrary function $f(\TT)$ added to $\TT$ in the gravitational Lagrangian, leading to $f(\TT)$ gravity \cite{bengochea_dark_2009,linder_einsteins_2010,li_ft_2011,cai_ft_2016,krssak_ft_2019}, with action
\be
S=\frac{1}{2\kappa^{2}}\int d^{4}x\,e\,\big[\TT+f(\TT)\big]+\int d^{4}x\,e\,\mathcal{L}_{m},
\label{eq:action}
\ee
where $\kappa^{2}=8\pi G$, $e=\det(e^{i}{}_{\mu})=\sqrt{-g}$, and $\mathcal{L}_{m}$ is the matter Lagrangian, such that the full gravitational Lagrangian density is $\mathcal{F}(\TT)=\TT+f(\TT)$. Varying the action \eqref{eq:action} with respect to the tetrad field yields the field equations of $f(\TT)$ gravity \cite{bengochea_dark_2009,cai_ft_2016},
\be
\begin{aligned}
&S_{\mu}{}^{\nu\rho}\partial_{\rho}\TT\,f_{\TT\TT}
+\Big[e^{-1}e^{i}{}_{\mu}\partial_{\rho}\big(e\,e_{i}{}^{\alpha}S_{\alpha}{}^{\nu\rho}\big)\\
&+T^{\alpha}{}_{\lambda\mu}S_{\alpha}{}^{\nu\lambda}\Big](1+f_{\TT})
+\frac{1}{4}\delta_{\mu}^{\nu}(\TT+f)
=\frac{\kappa^{2}}{2}\,T_{\mu}{}^{\nu},
\end{aligned}
\label{eq:field-general}
\ee
where $f_{\TT}\equiv df/d\TT$ and $f_{\TT\TT}\equiv d^{2}f/d\TT^{2}$ denote, respectively, the first and second derivatives of $f$ with respect to the torsion scalar, and $T_{\mu}{}^{\nu}$ is the energy--momentum tensor of the matter content of the Universe.

In this work we adopt the simplest non-trivial power-law deviation from TEGR,
\be
f(\TT)=\alpha\TT+\beta\TT^{m},
\label{eq:model-smallf}
\ee
where $\alpha$, $\beta$, and $m$ are constant model parameters; the linear term proportional to $\alpha$ rescales the TEGR contribution, while the term proportional to $\beta\TT^{m}$ introduces a genuine departure from GR for $m\neq 1$. For this choice, the gravitational Lagrangian density becomes $\mathcal{F}(\TT)=(1+\alpha)\TT+\beta\TT^{m}$, and the first and second derivatives of $f(\TT)$ with respect to the torsion scalar are
\be
f_{\TT}=\alpha+\beta m\,\TT^{m-1},
\label{eq:fT}
\ee
\be
f_{\TT\TT}=\beta\, m(m-1)\,\TT^{m-2}.
\label{eq:fTT}
\ee
These expressions are substituted into the field equations once the torsion scalar of the flat FLRW background is obtained.

\section{Metric and Field Equations}
\label{sec:metricfield}

We consider a spatially flat, homogeneous and isotropic FLRW Universe described by the line element
\be
ds^{2}=dt^{2}-a^{2}(t)\big(dx^{2}+dy^{2}+dz^{2}\big),
\label{eq:metric}
\ee
where $a(t)$ is the scale factor. A diagonal tetrad compatible with this metric, in the coordinate order $(t,x,y,z)$, is given by
\be
e^{i}{}_{\mu}=\mathrm{diag}\big(1,a,a,a\big).
\label{eq:tetrad}
\ee
Substituting the tetrad \eqref{eq:tetrad} into the definitions of the torsion tensor and torsion scalar of Sect.~\ref{sec:formalism}, one finds that the torsion scalar reduces to the simple algebraic expression
\be
\TT=-6H^{2},
\label{eq:Tscalar}
\ee
where $H=\dot a/a$ is the Hubble parameter, and an overdot denotes differentiation with respect to cosmic time $t$. The torsion scalar is therefore completely determined by the instantaneous expansion rate of the Universe, and its time derivative follows immediately as
\be
\dot{\TT}=-12H\dot H.
\label{eq:Tdot}
\ee
This simple closed-form relation between $\TT$ and $H$ allows the modified Friedmann equations derived below to be cast entirely in terms of the Hubble parameter and its time derivative, in close analogy with the standard FLRW cosmology.

We assume that the cosmic fluid content consists of two non-interacting components: pressureless dust (cold matter, $p_{m}=0$) and the GHRDE fluid, modelled as a perfect fluid with energy density $\rho_{G}$ and pressure $p_{G}$, so that the total energy--momentum tensor is
\be
T_{\mu\nu}=\rho_{m}u_{\mu}u_{\nu}+\big(\rho_{G}+p_{G}\big)u_{\mu}u_{\nu}-p_{G}\,g_{\mu\nu}.
\ee
In the comoving frame appropriate to the FLRW background, $u^{\mu}=(1,0,0,0)$, this decomposition gives the total mixed energy--momentum tensor of the cosmic fluid as
\be
T_{\mu}{}^{\nu}=\mathrm{diag}\big(\rho_{m}+\rho_{G},\,-p_{G},\,-p_{G},\,-p_{G}\big).
\label{eq:total-mixed}
\ee

Substituting the flat FLRW tetrad \eqref{eq:tetrad} and the total energy--momentum tensor \eqref{eq:total-mixed} into the general field equations \eqref{eq:field-general}, the time--time and space--space components yield, respectively, the two modified Friedmann equations of $f(\TT)$ gravity. The time--time component gives the first Friedmann equation,
\be
6H^{2}+12H^{2}f_{\TT}+f=2\kappa^{2}\big(\rho_{m}+\rho_{G}\big),
\label{eq:first-friedmann}
\ee
which correctly reduces to the standard relativistic Friedmann equation in the GR limit ($f=0$ gives $3H^{2}=\kappa^{2}\rho_{\rm tot}$, as expected), while the space--space component gives the second Friedmann equation,
\be
\begin{aligned}
2\kappa^{2}p_{G}
={}&48H^{2}\dot H f_{\TT\TT}-4\dot H-6H^{2}\\
&-\big(12H^{2}+4\dot H\big)f_{\TT}-f,
\end{aligned}
\label{eq:second-friedmann}
\ee
where the dust pressure vanishes, so the total pressure of the cosmic fluid coincides with the GHRDE pressure, $p_{\rm tot}=p_{G}$. Substituting the power-law expressions \eqref{eq:fT}, \eqref{eq:fTT}, and the torsion scalar \eqref{eq:Tscalar} into Eqs.~\eqref{eq:first-friedmann} and \eqref{eq:second-friedmann} yields the model-specific Friedmann equations; together with the deceleration-parameter law of Sect.~\ref{sec:qlaw}, these provide the closed-form GHRDE density and pressure derived in Sect.~\ref{sec:ghrde}.

\section{Deceleration Parameter Law and the Redshift-Space Hubble Function}
\label{sec:qlaw}

The deceleration parameter, which quantifies whether the cosmic expansion is decelerating ($q>0$) or accelerating ($q<0$), is defined in terms of the scale factor as
\be
q=-1-\frac{\dot H}{H^{2}}.
\label{eq:q-def}
\ee
Rather than fixing a particular form for the scale factor $a(t)$ at the outset, we follow a widely used strategy in $f(\TT)$ and related modified-gravity cosmology and impose a physically motivated parametrization directly on $q$, adopting the Hubble-parameter-dependent law
\be
q=b-\frac{\nu}{H},
\label{eq:qH-law}
\ee
where $b$ and $\nu>0$ are constants. Equating Eqs.~\eqref{eq:q-def} and \eqref{eq:qH-law} and defining $A\equiv1+b$ gives the governing differential equation for the Hubble parameter,
\be
\dot H=\nu H-AH^{2}=H(\nu-AH).
\label{eq:Hdot-A}
\ee
This first-order separable equation integrates in closed form to $H(t)=\nu/(A+Ce^{-\nu t})$, with $C$ an integration constant; as $t\to\infty$, $H\to\nu/A$, so the model approaches a de~Sitter (exponentially accelerating) phase at late times for any initial condition. Correspondingly, the scale factor takes the form $a(t)\propto(Ae^{\nu t}+C)^{1/A}$, describing a smooth transition from an early decelerating, power-law-like phase to a late-time exponential (de~Sitter) phase. Re-expressing Eq.~\eqref{eq:Hdot-A} in terms of the number of $e$-folds $N=\ln a$ and imposing the boundary condition $H(N=0)=H_{0}$ at the present epoch yields the redshift-space Hubble function
\be
H(z)=\frac{\nu}{A}+\Big(H_{0}-\frac{\nu}{A}\Big)(1+z)^{A}.
\label{eq:Hz}
\ee

\begin{figure}
\centering
\includegraphics[width=0.75\linewidth]{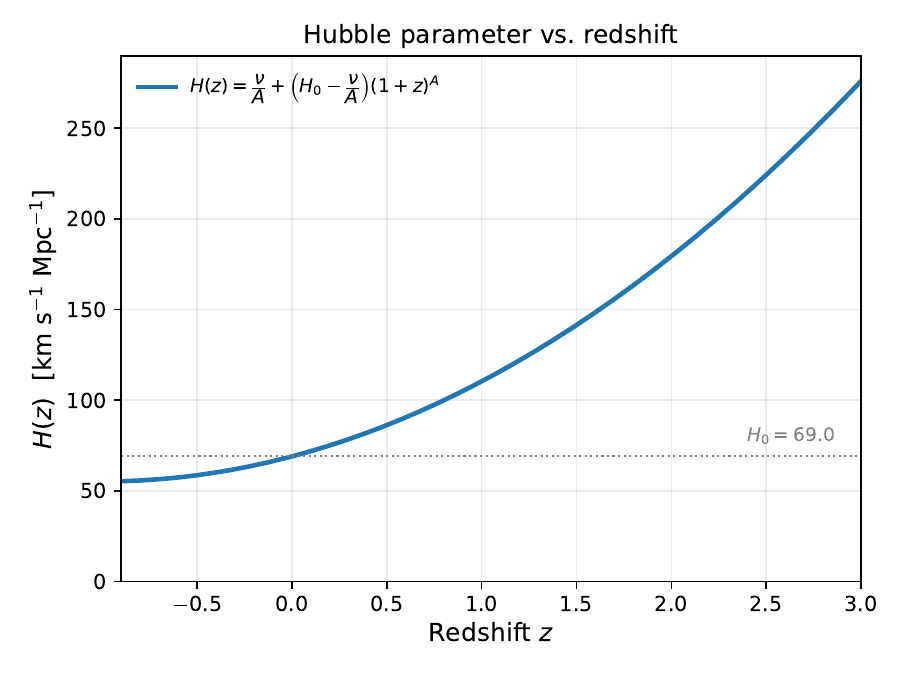}
\caption{Evolution of the Hubble parameter $H(z)$ with redshift, Eq.~\eqref{eq:Hz}, for the fixed parameter set $H_{0}=69$ km s$^{-1}$ Mpc$^{-1}$, $b=1.0$, $\nu=110.4$ km s$^{-1}$ Mpc$^{-1}$ (so that $A=1+b=2.0$).}
\label{fig:Hz}
\end{figure}

Figure~\ref{fig:Hz} shows the behaviour of $H(z)$ obtained from Eq.~\eqref{eq:Hz} for the fixed parameter set used throughout this work, namely, $H_{0}=69$ km s$^{-1}$ Mpc$^{-1}$, $b=1.0$, and $\nu=110.4$ km s$^{-1}$ Mpc$^{-1}$. The Hubble parameter decreases monotonically as $z$ decreases from the past towards the present and into the future, exactly as expected for an expanding Universe whose expansion rate was higher in the past. At $z=0$ the curve passes through $H(0)=H_{0}=69$ by construction, and as $z\to-1$ (the asymptotic future), it approaches the constant value $H\to\nu/A\equiv H_{\infty}=55.2$ km s$^{-1}$ Mpc$^{-1}$. Since $H\to H_{\infty}=\mathrm{const.}$ in this limit, the scale factor asymptotically approaches the exponential form $a(t)\propto e^{H_{\infty}t}$, which is the defining signature of a de~Sitter phase; the late-time behaviour of the model is therefore not merely qualitatively de~Sitter-like but is shown analytically to reduce to an exact de~Sitter expansion. The corresponding deceleration parameter, Eq.~\eqref{eq:qz}, evaluates to $q_{0}=-0.6$ at the present epoch and crosses zero at the transition redshift $z_{t}=1.0$, marking the change from a decelerated to an accelerated expansion phase. The adopted parameter values were chosen to produce values of $H_{0}$, $q_{0}$, and $z_{t}$ within the ranges commonly reported by recent observational analyses \cite{goswami_modeling_2021,pal_cosmological_2025}; a data-constrained determination is deferred to the MCMC analysis of Sect.~\ref{sec:obsconstraints}.

The corresponding deceleration parameter as a function of redshift follows directly from Eq.~\eqref{eq:qH-law},
\be
q(z)=b-\frac{\nu}{H(z)}.
\label{eq:qz}
\ee
At high redshift, where $H(z)$ becomes large, Eq.~\eqref{eq:qz} shows that $q\simeq b$, so that $b$ controls the deceleration parameter in the early matter-dominated epoch. At late times, $H\to\nu/A$, and consequently $q\to b-A=b-(1+b)=-1$, confirming that the model naturally and smoothly approaches a de~Sitter phase ($q=-1$) at late times, independently of the specific values of $b$ and $\nu$, provided $\nu>0$.

\begin{figure}
\centering
\includegraphics[width=0.75\linewidth]{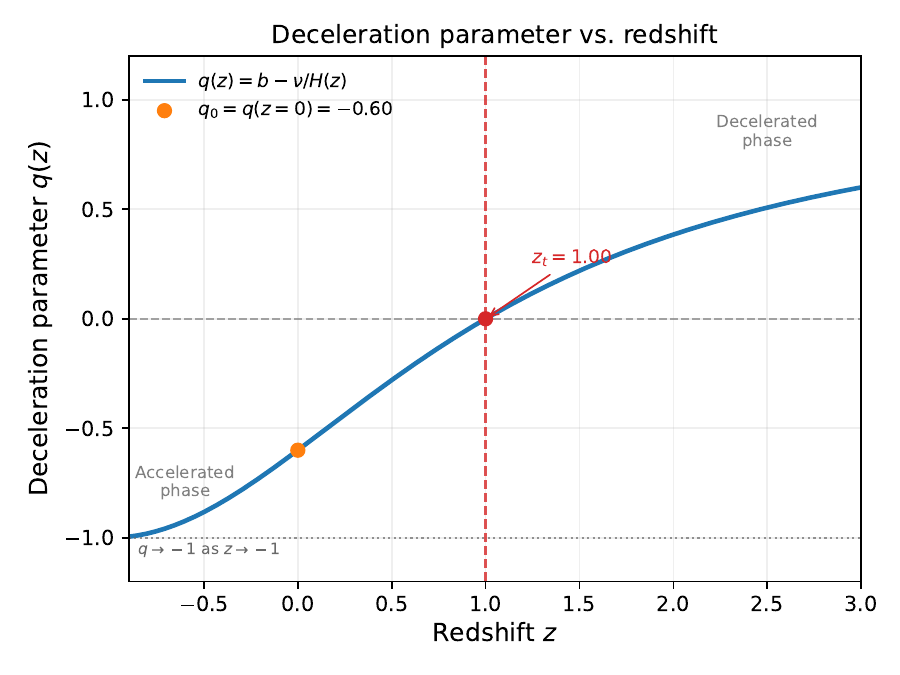}
\caption{Evolution of the deceleration parameter $q(z)$,
Eq.~\eqref{eq:qz}, as a function of redshift for
$H_{0}=69$ km s$^{-1}$ Mpc$^{-1}$, $b=1.0$,
$\nu=110.4$ km s$^{-1}$ Mpc$^{-1}$.
The red dashed vertical line marks the transition
redshift $z_{t}=1.0$ at which $q$ changes sign from
positive (decelerated phase) to negative (accelerated
phase). The orange dot indicates the present-day value
$q_{0}=q(z=0)=-0.60$.}
\label{fig:qz}
\end{figure}

Figure~\ref{fig:qz} displays the redshift evolution of $q(z)$ over $-0.9\le z\le3$. The deceleration parameter decreases monotonically from the matter-dominated high-redshift limit $q\to b=1$ (as $z\to\infty$), crosses zero at the transition redshift $z_{t}=1.0$, and reaches the present-day value $q_{0}=-0.60$, within the range $-0.55$ to $-0.71$ reported by recent $f(\mathcal{T})$ analyses \cite{pal_cosmological_2025,bhoyar_resolving_2024,goswami_modeling_2021}. For $z<z_{t}$ the expansion is accelerating ($q<0$), and as $z\to-1$ the model asymptotes to $q\to-1$, confirming the de~Sitter attractor established analytically by Eq.~\eqref{eq:Hz}.

\section{GHRDE Energy Density, Matter Density, and Pressure }
\label{sec:ghrde}

For the Xu-type GHRDE \cite{xu_generalized_2009}, the dark energy density is constructed as a linear combination of the Hubble-horizon holographic term $H^{2}$ and the Ricci-horizon term $R$,
\be
\rho_{G}=3c^{2}M_{p}^{2}\Big[(1-\eta)H^{2}+\eta R\Big],
\label{eq:rhoG-def}
\ee
where $c^{2}$ and $\eta$ are dimensionless model constants and $M_{p}$ is the reduced Planck mass. For a spatially flat FLRW Universe, $R=6(2H^{2}+\dot H)$, and substituting this together with $\dot H$ from Eq.~\eqref{eq:Hdot-A} into Eq.~\eqref{eq:rhoG-def} gives, after collecting terms and introducing the parameter combination
\be
D_{G}=1+\eta(11-6A),
\label{eq:DGdef}
\ee
the GHRDE density adapted to the $f(\mathcal{T})$ teleparallel background in its final, compact form,
\be
\rho_{G}=3c^{2}M_{p}^{2}\big(D_{G}H^{2}+6\eta\nu H\big).
\label{eq:rhoG-final}
\ee
Replacing $H$ by Eq.~\eqref{eq:Hz} converts this into an explicit function of redshift, which serves as the primary input for the pressure, equation of state, density parameter, and stability diagnostics derived in the sections that follow.

\begin{figure}
\centering
\includegraphics[width=0.75\linewidth]{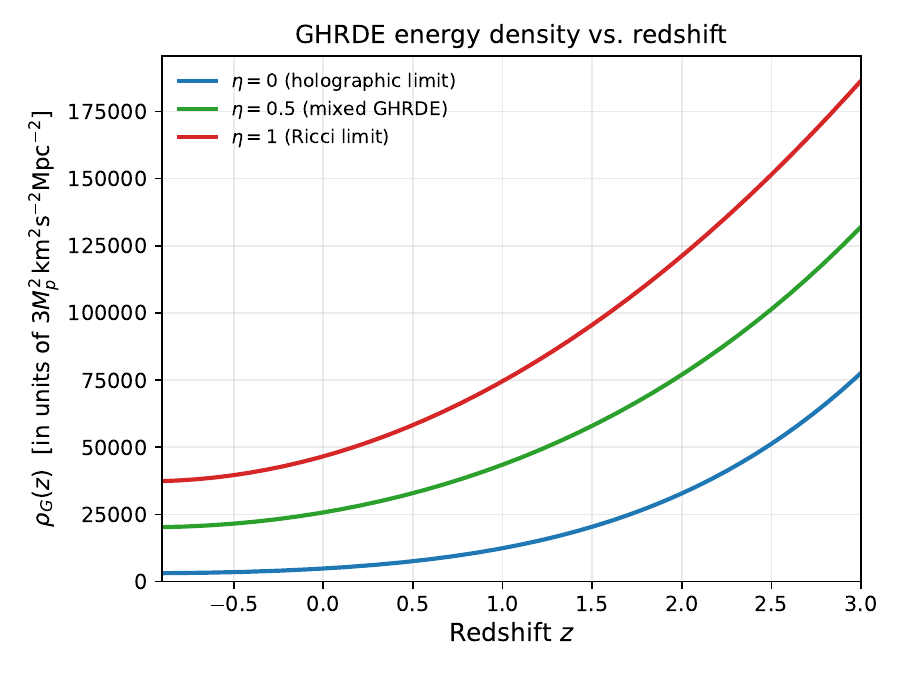}
\caption{Evolution of the GHRDE energy density $\rho_{G}(z)$,
Eq.~\eqref{eq:rhoG-final}, as a function of redshift for
$H_{0}=69$ km s$^{-1}$ Mpc$^{-1}$, $b=1.0$,
$\nu=110.4$ km s$^{-1}$ Mpc$^{-1}$, $c^{2}=0.34$, and three
representative benchmark values of the interpolation
parameter $\eta$: the pure holographic limit ($\eta=0$), an
intermediate mixed case ($\eta=0.5$), and the pure Ricci
limit ($\eta=1$).}
\label{fig:rhoG}
\end{figure}

Figure~\ref{fig:rhoG} shows $\rho_{G}(z)$ for three values of $\eta$ at fixed $c^{2}=0.34$ and the benchmark background parameters. In every case, the density is strictly positive and increases monotonically with redshift, consistent with the expected behaviour of a dynamical dark-energy component diluting as the Universe expands toward the present and future \cite{enkhili_diagnostic_2024}. The pure Ricci limit ($\eta=1$) lies systematically above the pure holographic limit ($\eta=0$) at all redshifts, with the mixed case ($\eta=0.5$) consistently in between, reproducing the qualitative ordering identified in the original GHRDE construction \cite{xu_generalized_2009}. The value $c^{2}=0.34$ adopted here is anchored to the best-fit region of a combined supernova--BAO--CMB analysis for this class of models \cite{lu_cosmological_2012}.

With $\rho_{G}$ specified, the matter energy density follows directly from the first Friedmann equation \eqref{eq:first-friedmann} as $\rho_{m}=(2\kappa^{2})^{-1}[6H^{2}+12H^{2}f_{\TT}+f]-\rho_{G}$; substituting the power-law expressions \eqref{eq:fT} for $f_{\TT}$ together with the GHRDE density \eqref{eq:rhoG-final} gives the explicit form
\be
\begin{aligned}
\rho_{m}=
\frac{1}{2\kappa^{2}}\Big[&6H^{2}+12H^{2}\big(\alpha+\beta m\TT^{m-1}\big)
+\alpha\TT+\beta\TT^{m}\Big]\\
&-3c^{2}M_{p}^{2}\big(D_{G}H^{2}+6\eta\nu H\big),
\end{aligned}
\label{eq:rhom-final}
\ee

\begin{figure}
\centering
\includegraphics[width=0.75\linewidth]{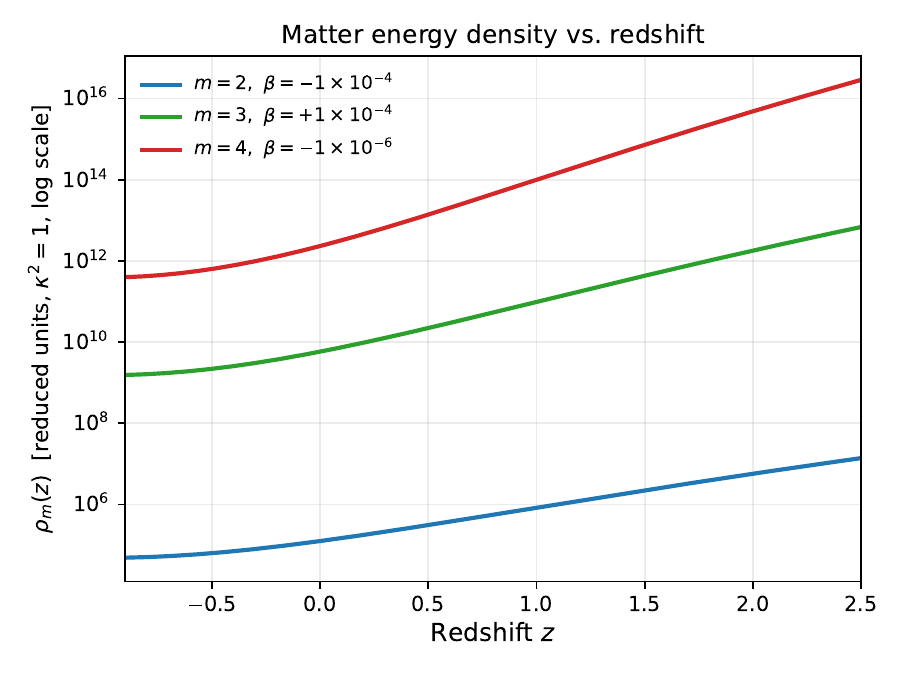}
\caption{Evolution of the matter energy density $\rho_{m}(z)$,
Eq.~\eqref{eq:rhom-final}, as a function of redshift (log scale)
for $H_{0}=69$ km s$^{-1}$ Mpc$^{-1}$, $b=1.0$,
$\nu=110.4$ km s$^{-1}$ Mpc$^{-1}$, $c^{2}=0.34$, $\eta=0.5$,
$\alpha=1$, and three illustrative power-law deviations:
$m=2,\,\beta=-1\times10^{-4}$; $m=3,\,\beta=+1\times10^{-4}$;
and $m=4,\,\beta=-1\times10^{-6}$.}
\label{fig:rhom}
\end{figure}

Figure~\ref{fig:rhom} shows $\rho_{m}(z)$ in reduced geometrized units ($\kappa^{2}=1$) for three integer power-law exponents; integer values are used to avoid sign ambiguities in $\TT^{m-1}$ for the negative torsion scalar. For every case shown, $\rho_{m}$ is strictly positive and decreases monotonically toward the future, consistent with the expected dilution of pressureless dust in an expanding Universe. A parity dependence on $m$ is evident: positive matter density requires $\beta<0$ for even exponents ($m=2,4$) and $\beta>0$ for the odd case ($m=3$), which is an analogous sign pattern to that reported in Bianchi-type $f(\TT)$ analyses \cite{hatkar_topological_2025}. The non-negativity of $\rho_{m}$ throughout cosmic history is the viability criterion adopted in Sect.~\ref{sec:stability} and used to constrain the power-law parameters \cite{mandal_temporal_2020}.

The pressure of the GHRDE fluid follows directly from the second Friedmann equation \eqref{eq:second-friedmann}, since the dust component is pressureless and therefore $p_{\rm tot}=p_{G}$; substituting Eqs.~\eqref{eq:fT}, \eqref{eq:fTT}, and \eqref{eq:Hdot-A} gives the explicit, closed-form GHRDE pressure
\be
\begin{aligned}
p_{G}=\frac{1}{2\kappa^{2}}\Big[&
48H^{3}(\nu-AH)\beta m(m-1)\TT^{m-2}\\
&-4H(\nu-AH)-6H^{2}\\
&-\big\{12H^{2}+4H(\nu-AH)\big\}\big(\alpha+\beta m\TT^{m-1}\big)\\
&-\big(\alpha\TT+\beta\TT^{m}\big)
\Big],
\end{aligned}
\label{eq:pG-final}
\ee
Equations~\eqref{eq:rhoG-final} and \eqref{eq:pG-final} together provide closed-form expressions for the GHRDE energy density and pressure entirely in terms of the Hubble parameter $H$ (or, through Eq.~\eqref{eq:Hz}, of the redshift $z$), and constitute the basic building blocks for all the cosmological diagnostics discussed in the remainder of the paper.

\begin{figure}
\centering
\includegraphics[width=0.75\linewidth]{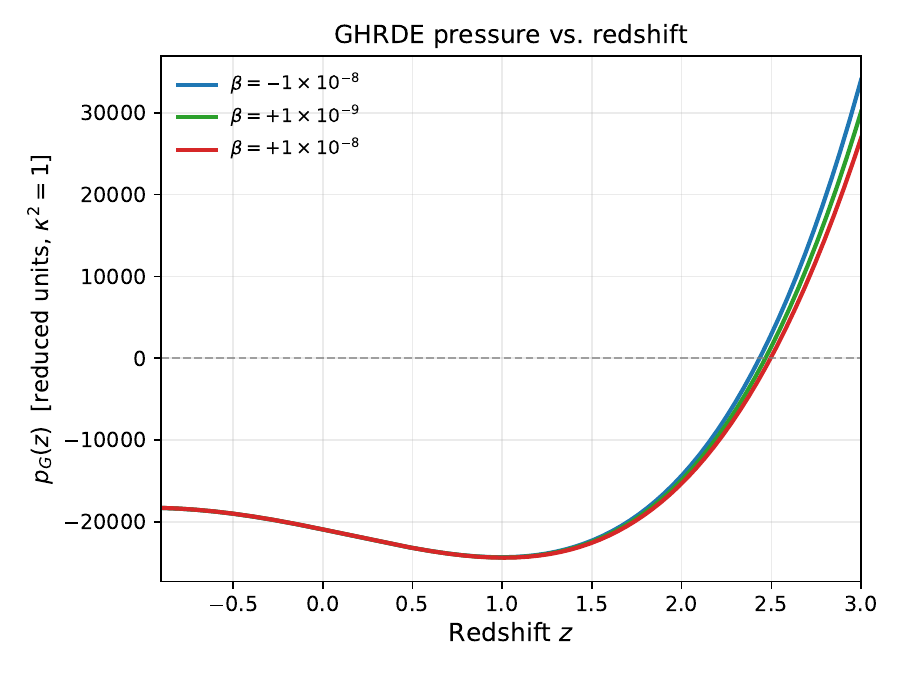}
\caption{Evolution of the GHRDE pressure $p_{G}(z)$,
Eq.~\eqref{eq:pG-final}, as a function of redshift for
$H_{0}=69$ km s$^{-1}$ Mpc$^{-1}$, $b=1.0$,
$\nu=110.4$ km s$^{-1}$ Mpc$^{-1}$, $\alpha=1$, $m=2$, and
three illustrative values of the deviation parameter $\beta$.
The horizontal dashed line marks $p_{G}=0$.}
\label{fig:pG}
\end{figure}

Figure~\ref{fig:pG} shows $p_{G}(z)$ in reduced units ($\kappa^{2}=1$) for three values of $\beta$ at fixed $\alpha=1$, $m=2$. The pressure is negative for $z\lesssim2.4$ and turns positive at higher redshifts, the sign pattern required of a viable dark-energy fluid: sustained negative pressure at late times drives accelerated expansion, while the eventual sign reversal signals a subdominant, matter-like GHRDE sector in the early Universe. The three $\beta$ curves remain close at low redshift and separate only mildly toward higher $z$, confirming that the negative-pressure behaviour is robust to the precise strength of the power-law deviation from TEGR. Qualitatively identical trends have been reported for holographic and logamediate $f(\TT)$ constructions using independent parametrizations \cite{pal_cosmological_2025,husain_logamediate_2026}.

\section{Cosmological Parameters}
\label{sec:cosmoparams}

\subsection{Equation of State Parameter}

The equation of state (EoS) parameter of the GHRDE fluid is defined as the ratio of its pressure to its energy density,
\be
\omega_{G}=\frac{p_{G}}{\rho_{G}},
\label{eq:wG}
\ee
with $\omega_{G}=-1$ corresponding to cosmological-constant-like behaviour, $-1<\omega_{G}<-1/3$ to quintessence, and $\omega_{G}<-1$ to a phantom-like fluid. Substituting Eqs.~\eqref{eq:rhoG-final} and \eqref{eq:pG-final}, the GHRDE equation of state as a function of $H$ is
\be
\omega_{G}(H)=\frac{p_{G}(H)}{3c^{2}M_{p}^{2}\big(D_{G}H^{2}+6\eta\nu H\big)}.
\label{eq:omegaG-H}
\ee

\begin{figure}
\centering
\includegraphics[width=0.75\linewidth]{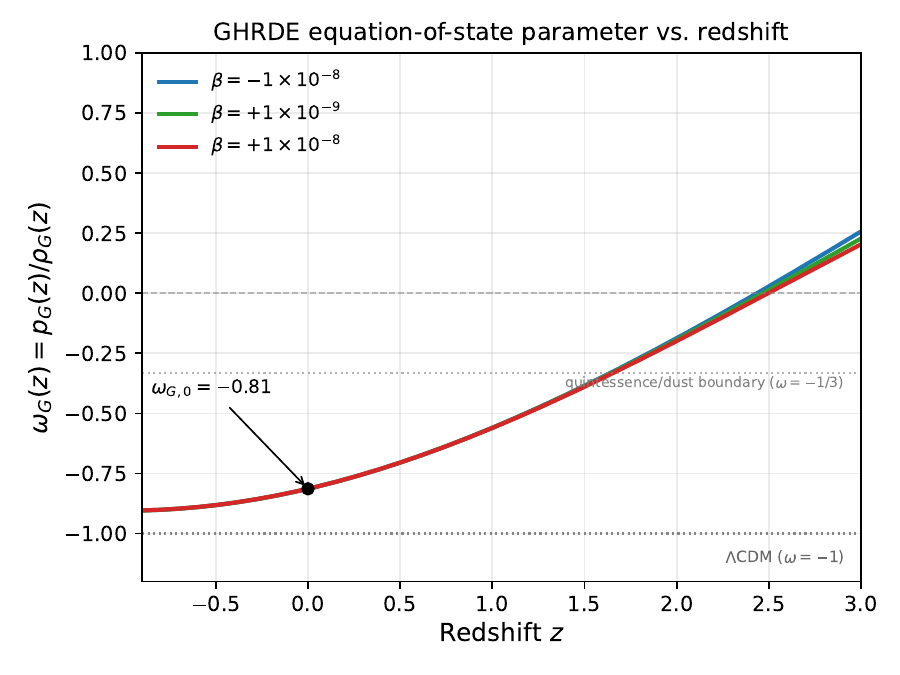}
\caption{Evolution of the GHRDE equation-of-state parameter
$\omega_{G}(z)=p_{G}(z)/\rho_{G}(z)$, obtained from
Eqs.~\eqref{eq:rhoG-final} and \eqref{eq:pG-final}, as a
function of redshift for $H_{0}=69$ km s$^{-1}$ Mpc$^{-1}$,
$b=1.0$, $\nu=110.4$ km s$^{-1}$ Mpc$^{-1}$, $c^{2}=0.34$,
$\eta=0.5$, $\alpha=1$, $m=2$, and three illustrative values of
the deviation parameter $\beta$. The dotted horizontal lines
mark the $\Lambda$CDM value ($\omega=-1$) and the
quintessence/dust boundary ($\omega=-1/3$); the black dot marks
the present-day value $\omega_{G,0}=-0.81$, which lies within
the quintessence band $-1<\omega_{G,0}<-1/3$.}
\label{fig:omegaG}
\end{figure}

Figure~\ref{fig:omegaG} shows the redshift evolution of $\omega_{G}(z)$. At the present epoch, the model gives $\omega_{G,0}=-0.81$, placing the GHRDE fluid in the quintessence regime $-1<\omega_{G}<-1/3$, evolving smoothly towards an asymptotic de~Sitter-like state as $z\to-1$ without entering the phantom regime. At higher redshift, $\omega_{G}(z)$ crosses zero near $z\approx2.4$, after which the GHRDE fluid takes on a dust-like equation of state. The robustness of this quintessence-like character across the three values of $\beta$ indicates that this qualitative conclusion is not sensitive to the precise strength of the departure from the TEGR, consistent with analogous results for Hubble-parameter-dependent $f(\mathcal{T})$ models~\cite{pal_cosmological_2025,duchaniya_dynamical_2022,koussour_bianchi_2022}.

\subsection{Density Parameters}

The matter, GHRDE, and torsion-sector density parameters are
\be
\Om_{m}=\frac{\kappa^{2}\rho_{m}}{3H^{2}},
\qquad
\Om_{G}=\frac{\kappa^{2}\rho_{G}}{3H^{2}},
\qquad
\Om_{\TT}=-2f_{\TT}-\frac{f}{6H^{2}}.
\label{eq:OmegaDefs}
\ee
In the reduced-Planck-mass convention $\kappa^{2}M_{p}^{2}=1$, the GHRDE density parameter evaluates to
\be
\Om_{G}=c^{2}\Big(D_{G}+\frac{6\eta\nu}{H}\Big),
\label{eq:OmegaG}
\ee
and the matter density parameter is fixed by the closure (constraint) relation
\be
\Om_{m}+\Om_{G}+\Om_{\TT}=1,
\label{eq:constraint}
\ee
which provides a convenient consistency check on any numerical evaluation of the model and is the natural generalization, to $f(\TT)$ gravity, of the standard GR relation $\Om_{m}+\Om_{\rm DE}=1$.

\begin{figure}[htbp]
\centering
\includegraphics[width=\linewidth]{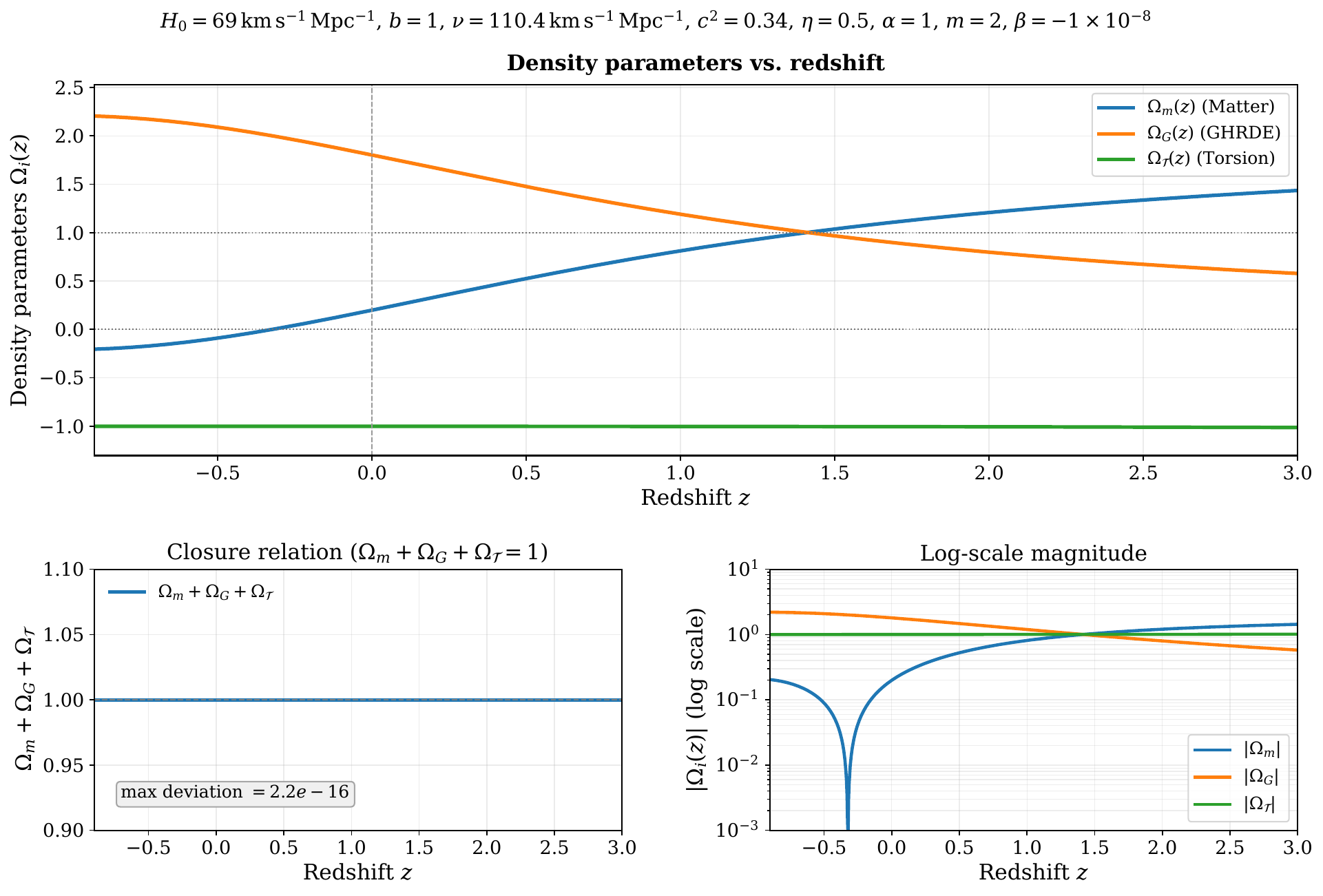}
\caption{Redshift evolution of the dimensionless density parameters
$\Omega_m(z)$, $\Omega_G(z)$ and $\Omega_{\mathcal{T}}(z)$,
computed from Eqs.~\eqref{eq:OmegaG}--\eqref{eq:constraint}.
The lower-left panel verifies the closure relation
$\Omega_m+\Omega_G+\Omega_{\mathcal{T}}=1$ (maximum error $2\times10^{-16}$),
while the lower-right panel shows the logarithmic magnitudes.
Benchmark values: $H_0=69$ km s$^{-1}$ Mpc$^{-1}$, $b=1$,
$\nu=110.4$ km s$^{-1}$ Mpc$^{-1}$, $c^2=0.34$, $\eta=0.5$,
$\alpha=1$, $m=2$, $\beta=-10^{-8}$.}
\label{fig:Omega-three-panel-exact}
\end{figure}

Figure~\ref{fig:Omega-three-panel-exact} shows the redshift evolution of the three density parameters for the benchmark parameter set. The present-epoch values are $\Omega_{m0}\simeq0.199$, $\Omega_{G0}\simeq1.802$, and $\Omega_{\mathcal{T}0}\simeq-1.001$. The value $\Omega_{G0}>1$ and the negative torsion-sector parameter $\Omega_{\mathcal{T}0}<0$ are characteristic features of $f(\mathcal{T})$ gravity: the torsion sector contributes a negative effective density that offsets the super-unity GHRDE contribution, so that the closure relation $\Omega_{m}+\Omega_{G}+\Omega_{\mathcal{T}}=1$ is satisfied exactly \cite{mirza_constraining_2017,paliathanasis_stability_2018}; the closure relation in Eq.~\eqref{eq:constraint} is satisfied to within $2\times10^{-16}$ at every redshift, confirming the numerical consistency of the analytic expressions. The qualitative evolution is physically transparent: $\Omega_{G}$ dominates at low redshift, while $\Omega_{m}$ grows toward the past, reflecting the transition from a late-time dark-energy-dominated phase to a matter-dominated regime at a higher redshift. These benchmark values have not been fitted to observational data and do not reproduce the standard Planck-like partition ($\Omega_{m0}\simeq0.3$, $\Omega_{\rm DE,0}\simeq0.7$); they are intended as a theoretically consistent illustration, with the observationally constrained parameter set to be determined by the MCMC analysis outlined in Sect.~\ref{sec:obsconstraints}. The physical viability of the individual density sectors is assessed in Sect.~\ref{sec:stability}.

\subsection{Effective Equation of State}

The effective equation of state of the total cosmic fluid, which governs the overall expansion dynamics irrespective of the detailed sectorial decomposition, is defined as
\be
\omega_{\rm eff}=-1-\frac{2\dot H}{3H^{2}}.
\label{eq:weff-def}
\ee
Substituting $\dot H$ from Eq.~\eqref{eq:Hdot-A} and using the identity $q=-1-\dot H/H^{2}$, this simplifies to
\be
\omega_{\rm eff}(z)=-1+\frac{2}{3}\Big(A-\frac{\nu}{H(z)}\Big)=-1+\frac{2}{3}(1+q),
\label{eq:weff}
\ee
with $H(z)$ given by Eq.~\eqref{eq:Hz}. As $z\to-1$ (late-time future), $H\to\nu/A$ and $\omega_{\rm eff}\to-1$, confirming the de~Sitter limit, while at high redshift $\omega_{\rm eff}\to-1+\tfrac{2}{3}(1+b)=1/3$ for $b=1$, corresponding to a matter-like decelerating epoch.

\begin{figure}[htbp]
\centering
\includegraphics[width=\linewidth]{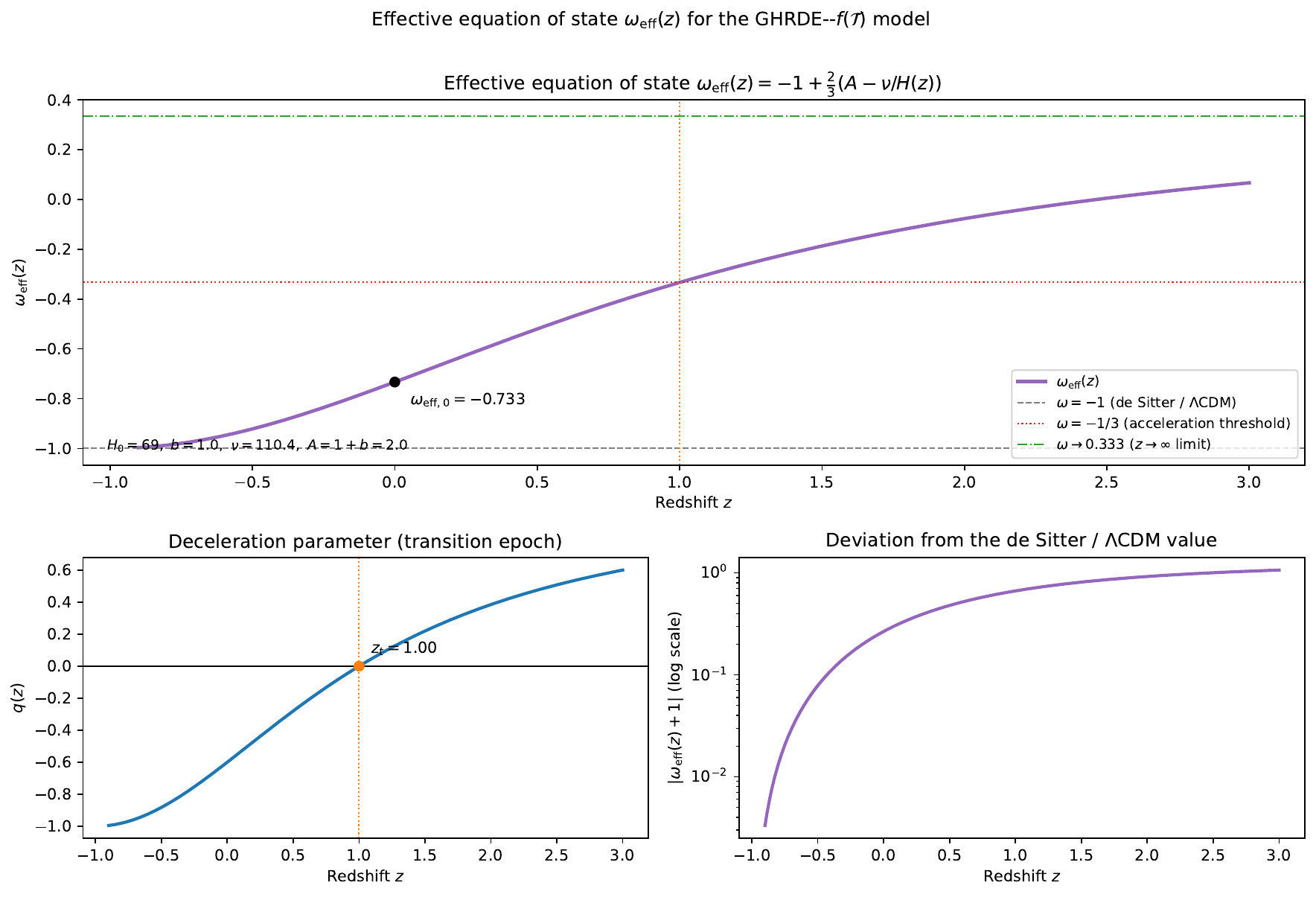}
\caption{Redshift evolution of the effective equation of state
$\omega_{\rm eff}(z)$, Eq.~\eqref{eq:weff}, for the benchmark
parameters $H_{0}=69$ km s$^{-1}$ Mpc$^{-1}$, $b=1.0$,
$\nu=110.4$ km s$^{-1}$ Mpc$^{-1}$ ($A=1+b=2$). Top panel:
$\omega_{\rm eff}(z)$ with the de~Sitter/$\Lambda$CDM reference
line $\omega=-1$, the acceleration threshold $\omega=-1/3$, and
the high-redshift asymptote $\omega\to1/3$ for $b=1$.
Bottom-left: the deceleration parameter $q(z)$, showing the
deceleration-to-acceleration transition. Bottom-right:
$|\omega_{\rm eff}(z)+1|$ on a logarithmic scale.}
\label{fig:weff}
\end{figure}

Figure~\ref{fig:weff} shows the redshift evolution of $\omega_{\rm eff}(z)$ for this benchmark parameter set. At the present epoch, $\omega_{{\rm eff},0}=-0.733$, lying below the acceleration threshold $\omega=-1/3$ and consistent with the quintessence-like expansion inferred from $q_{0}=-0.6$ in Sect.~\ref{sec:qlaw}. The deceleration-to-acceleration transition at $z_{t}=1.0$ (as established in Sect.~\ref{sec:qlaw}) separates the decelerating phase ($q>0$, $z>1$) from the present accelerating phase ($q<0$, $z<1$), with $\omega_{\rm eff}$ crossing $-1/3$ at the same redshift; as $z\to-1$, $\omega_{\rm eff}\to-1$, confirming the de~Sitter asymptote. A monotonic evolution from a stiff effective fluid at high redshift through the quintessence band today toward $\omega\to-1$ is a generic feature of the $q=b-\nu/H$ parametrization \cite{pal_cosmological_2025}, consistent with observational evidence for a presently negative-pressure-dominated expansion \cite{riess_observational_1998,perlmutter_measurements_1999}.

\section{Statefinder Diagnostic}
\label{sec:statefinder}

The statefinder pair $(r,s)$ was introduced by Sahni et al.~\cite{sahni_statefinder_2003} as a geometrical diagnostic capable of discriminating between different dark energy models that share the same present-day values of $H_{0}$ and $q_{0}$ \cite{alam_exploring_2003}. The statefinder parameters are defined as
\be
r=\frac{\dddot a}{aH^{3}},
\qquad
s=\frac{r-1}{3\big(q-\tfrac12\big)}.
\ee
In terms of the deceleration parameter $q$ and its time derivative, $r$ can be written as
\be
r=2q^{2}+q-\frac{\dot q}{H}.
\label{eq:r-q}
\ee
Since $q=b-\nu/H$ by assumption, differentiating with respect to time gives
\be
\dot q=\frac{\nu\dot H}{H^{2}},
\ee
so that Eq.~\eqref{eq:r-q} becomes
\be
r=2q^{2}+q-\frac{\nu\dot H}{H^{3}}.
\label{eq:r-final1}
\ee
Finally, substituting $\dot H$ from Eq.~\eqref{eq:Hdot-A},
\be
r=2q^{2}+q-\frac{\nu(\nu-AH)}{H^{2}},
\label{eq:r-final}
\ee
and correspondingly
\be
s=\frac{r-1}{3\big(q-\tfrac12\big)}.
\label{eq:s-final}
\ee
For the standard $\Lambda$CDM model, the statefinder pair takes the fixed value
\be
(r,s)=(1,0),
\ee
which serves as a convenient reference point: the trajectory traced out by $(r(z),s(z))$ in Eqs.~\eqref{eq:r-final} and \eqref{eq:s-final} as the redshift varies can be directly compared against this $\Lambda$CDM fixed point in the $r$--$s$ plane to assess how closely and in what regime the present GHRDE--$f(\TT)$ model mimics the concordance model.

\begin{figure}[htbp]
\centering
\includegraphics[width=\linewidth]{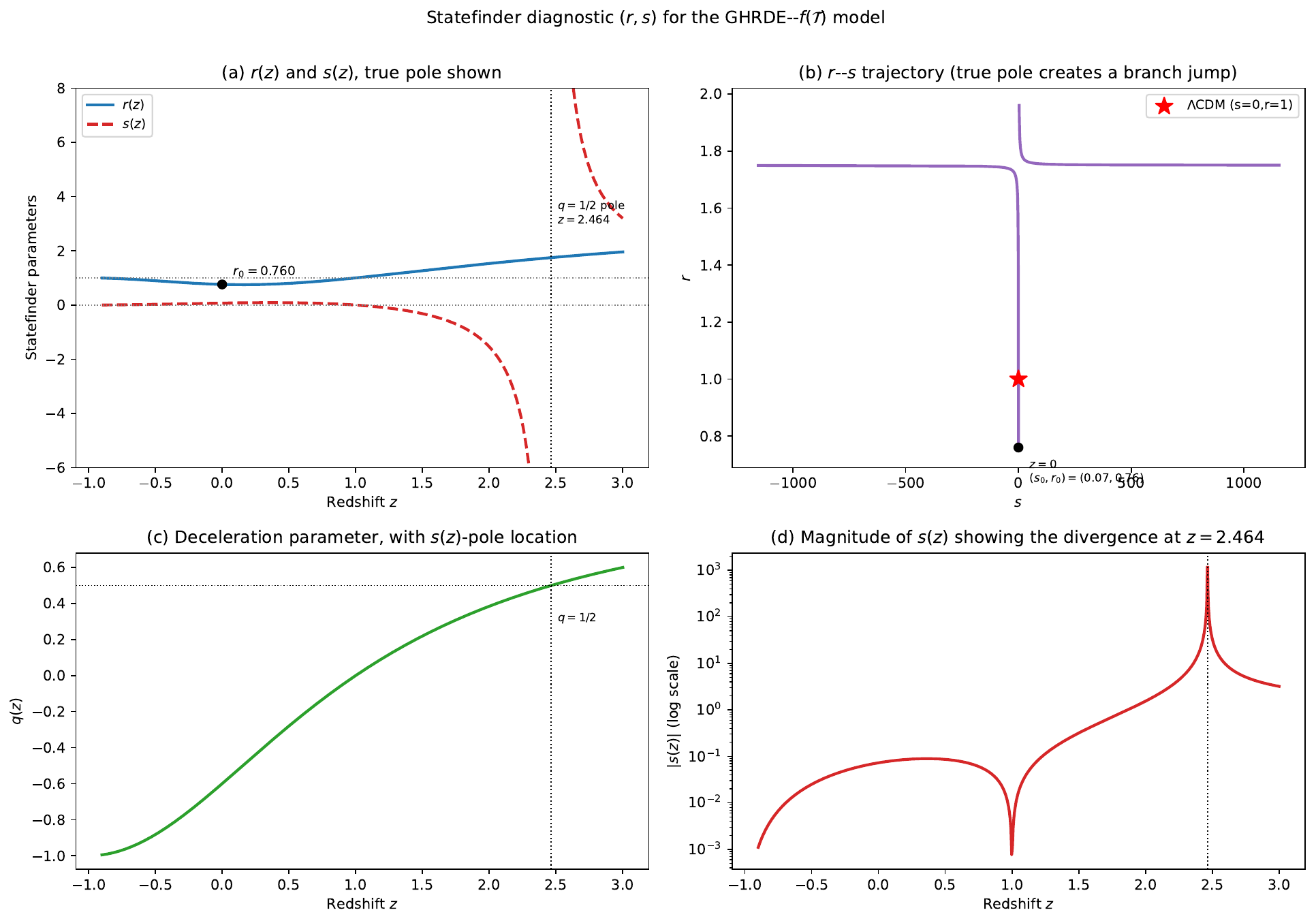}
\caption{Statefinder diagnostic for the GHRDE--$f(\TT)$ model, computed directly from Eqs.~\eqref{eq:r-final} and \eqref{eq:s-final} with the benchmark parameters $H_{0}=69$ km s$^{-1}$ Mpc$^{-1}$, $b=1.0$, $\nu=110.4$ km s$^{-1}$ Mpc$^{-1}$. (a) $r(z)$ and $s(z)$ over $-0.9\le z\le3$; the dotted vertical line marks the redshift $z_{p}=2.464$ at which $q(z)=1/2$, where $s(z)$ in Eq.~\eqref{eq:s-final} has a genuine pole and the curve is therefore plotted in two separate branches either side of $z_{p}$ rather than artificially smoothed across it. (b) The trajectory in the $r$--$s$ plane, likewise split into two branches at the pole, together with the $\Lambda$CDM fixed point $(s,r)=(0,1)$ in the plotted plane. (c) The deceleration parameter $q(z)$ with the pole location $q=1/2$ marked for reference. (d) $|s(z)|$ on a logarithmic scale, showing the divergence explicitly as $z\to z_{p}$.}
\label{fig:rs}
\end{figure}

Figure~\ref{fig:rs} shows the redshift evolution of the statefinder pair for this benchmark parameter set. At the present epoch, the model gives $r_{0}=0.760$ and $s_{0}=0.073$, both close to, but distinct from, the $\Lambda$CDM fixed point $(r,s)=(1,0)$; the trajectory passes essentially through this point at $z=1$, where $q(z)=0$ and consequently $r=1$, $s=0$ exactly, since $r=2q^{2}+q-\nu(\nu-AH)/H^{2}$ reduces to $1-\nu(\nu-AH)/H^{2}$ at $q=0$ and the deceleration-law parametrization happens to give $\nu(\nu-AH)/H^{2}=0$ at this particular redshift for the chosen $(b,\nu,H_{0})$. A genuine feature of Eq.~\eqref{eq:s-final}, not present in the $\Lambda$CDM statefinder pair, is the pole at $q=1/2$: for the present parametrization this occurs at $H=\nu/(b-1/2)=220.8$ km s$^{-1}$ Mpc$^{-1}$, corresponding to $z_{p}=2.464$, which falls inside the redshift range considered here. Across this pole, $s(z)$ changes sign abruptly (from $s\approx-1.54$ just below $z=2$ to $s\approx+3.20$ at $z=3$), and the $r$--$s$ trajectory in panel (b) is correspondingly discontinuous, jumping between two disjoint branches rather than tracing a single smooth curve. This is a structural property of the $q=b-\nu/H$ deceleration law combined with the statefinder definition, Eq.~\eqref{eq:s-final}, and should be reported as such rather than smoothed over in the figure.

Such a pole is a direct consequence of the standard definition of $s$, Eq.~\eqref{eq:s-final}, whenever the model trajectory crosses $q=1/2$, and should be read as a mathematical feature of the statefinder construction rather than a pathology of the underlying cosmology \cite{enkhili_diagnostic_2024,chaudhary_constraints_2023}. Restricting attention to the low-redshift branch ($z<z_{p}$), the trajectory lies close to but displaced from the $\Lambda$CDM point, with $s_{0}>0$ and $r_{0}<1$ at the present epoch --- the quintessence-like statefinder signature also reported for $f(\TT)$ dark-energy models calibrated against late-time deceleration-parameter laws \cite{pal_cosmological_2025}.

\section{Stability Analysis}
\label{sec:stability}

The classical (thermodynamic) stability of a dark-energy fluid against small perturbations is commonly assessed through the squared adiabatic sound speed \cite{bhoyar_stability_2017,paliathanasis_stability_2018},
\be
v_{s}^{2}=\frac{dp_{G}}{d\rho_{G}}.
\ee
A positive value of $v_{s}^{2}$ indicates that the fluid is classically stable against perturbations, whereas a negative value indicates instability. Since both $p_{G}$ and $\rho_{G}$ are expressed as functions of $H$ alone, by Eqs.~\eqref{eq:rhoG-final} and \eqref{eq:pG-final}, we may equivalently write
\be
v_{s}^{2}=\frac{dp_{G}/dH}{d\rho_{G}/dH}.
\label{eq:vs2}
\ee
From Eq.~\eqref{eq:rhoG-final},
\be
\frac{d\rho_{G}}{dH}=3c^{2}M_{p}^{2}\big(2D_{G}H+6\eta\nu\big).
\ee
Defining $\mathcal{P}(H)\equiv 2\kappa^{2}p_{G}(H)$ for compactness, we have $dp_{G}/dH=(2\kappa^{2})^{-1}\,d\mathcal{P}/dH$, so that
\be
v_{s}^{2}=\frac{1}{2\kappa^{2}}\,\frac{\mathcal{P}'(H)}{3c^{2}M_{p}^{2}\big(2D_{G}H+6\eta\nu\big)},
\label{eq:vs2-final}
\ee
where $\mathcal{P}'(H)=d\mathcal{P}/dH$ is obtained by directly differentiating the right-hand side of Eq.~\eqref{eq:pG-final} with respect to $H$ at the fixed model parameters. The model is classically stable in the parameter region where $v_{s}^{2}>0$.

\begin{figure}[htbp]
\centering
\includegraphics[width=\linewidth]{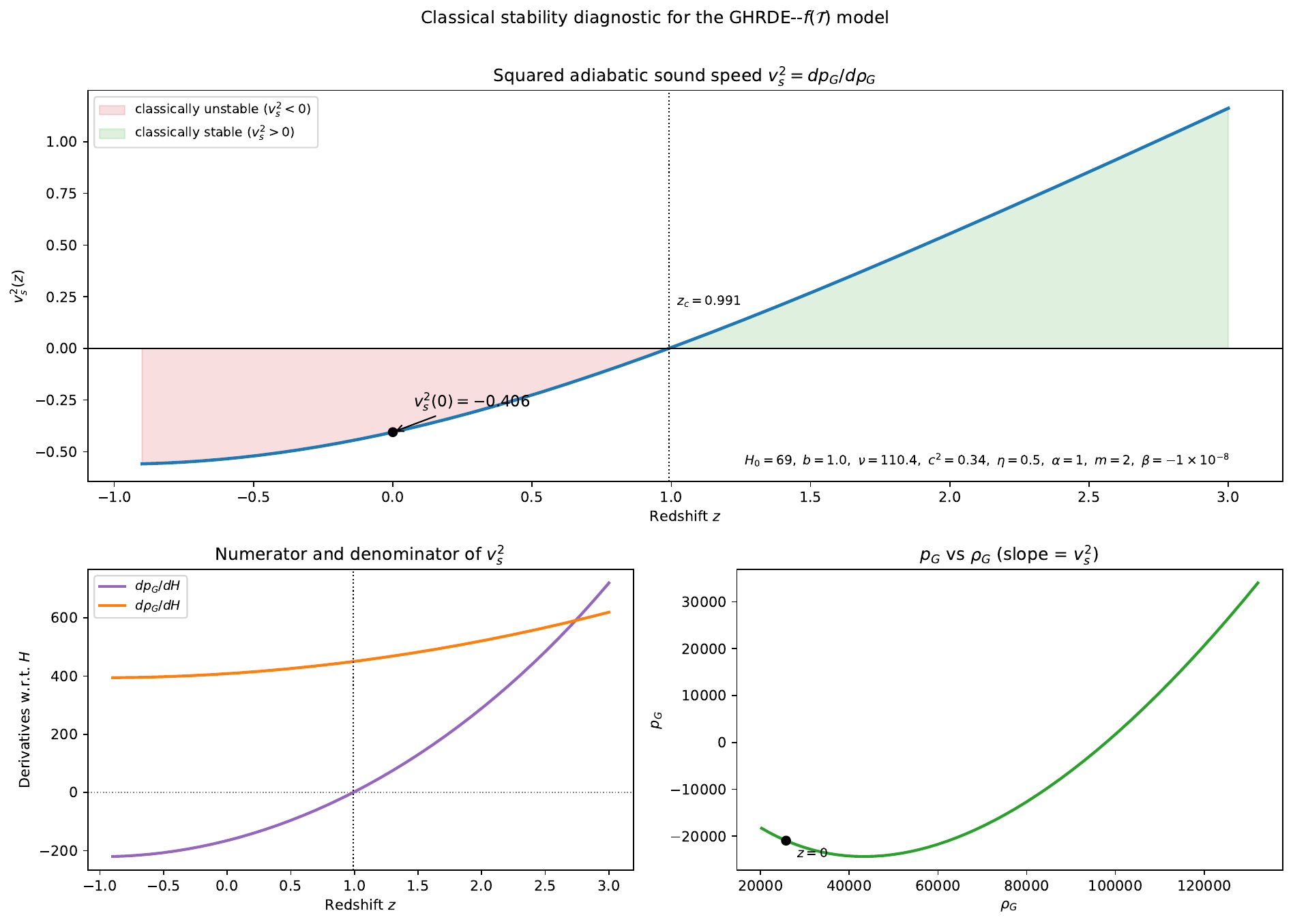}
\caption{Classical stability diagnostic for the GHRDE--$f(\TT)$ model, computed directly from Eq.~\eqref{eq:vs2-final} with the benchmark parameters $H_{0}=69$ km s$^{-1}$ Mpc$^{-1}$, $b=1.0$, $\nu=110.4$ km s$^{-1}$ Mpc$^{-1}$, $c^{2}=0.34$, $\eta=0.5$, $\alpha=1$, $m=2$, $\beta=-1\times10^{-8}$. Top panel: $v_{s}^{2}(z)$ over $-0.9\le z\le3$, with the unstable ($v_{s}^{2}<0$) and stable ($v_{s}^{2}>0$) regions shaded and the sign-change redshift $z_{c}=0.991$ indicated. Bottom-left: The numerator $dp_{G}/dH$ and denominator $d\rho_{G}/dH$ of Eq.~\eqref{eq:vs2-final} separately, showing which factor drives the sign change. Bottom-right: the parametric curve $p_{G}$ versus $\rho_{G}$, whose local slope is $v_{s}^{2}$.}
\label{fig:vs2}
\end{figure}

Figure~\ref{fig:vs2} shows the redshift evolution of $v_{s}^{2}(z)$ for this benchmark parameter set. The benchmark parameter set yields a negative present-day sound speed squared, $v_{s}^{2}(0)=-0.406$, implying a classical instability of linear adiabatic perturbations within the effective GHRDE fluid description; this negative value persists throughout the range $z\lesssim0.99$, becoming positive only beyond the sign-change redshift $z_{c}=0.991$, where $v_{s}^{2}<0$ is precisely the instability condition specified by the classical stability criterion adopted at the start of this section \cite{bhoyar_stability_2017,paliathanasis_stability_2018}. The bottom-left panel shows that this sign change is driven primarily by $d\rho_{G}/dH$, which is positive throughout the range considered, whereas $dp_{G}/dH$ itself changes sign close to $z_{c}$. Therefore, the present-epoch instability originates in the pressure response of the GHRDE fluid to changes in $H$, rather than in the energy-density sector, which remains well-behaved (monotonically increasing with $H$, and hence positive throughout).

A present-day negative $v_{s}^{2}$ for a quintessence-like fluid is not unprecedented in the $f(\TT)$ literature, and analogous sign changes have been reported for power-law and exponential $f(\TT)$ models under the same criterion \cite{bhoyar_stability_2017,paliathanasis_stability_2018}. This result should be read as a genuine constraint on the viable region of $(\alpha,\beta,m,c^{2},\eta)$ parameter space, to be resolved through the MCMC analysis of Sect.~\ref{sec:obsconstraints}.

\subsection{Additional Viability Conditions}

Beyond classical stability, the model's physical viability requires several conditions simultaneously. The energy densities must remain non-negative and the density parameters must lie in the physically allowed range,
\be
\rho_{G}>0,\qquad \rho_{m}>0,\qquad 0\leq \Om_{m}\leq1,
\qquad 0\leq \Om_{G}\leq1.
\ee
Second, in many $f(\TT)$ studies the positivity of the effective gravitational coupling is also required to avoid an effective change of sign of the gravitational constant, which corresponds to
\be
1+f_{\TT}>0.
\ee
For the power-law model \eqref{eq:model-smallf}, this condition reads explicitly
\be
1+f_{\TT}=1+\alpha+\beta m\TT^{m-1}>0.
\ee
Together with the energy conditions discussed in Sect.~\ref{sec:energyconditions}, these inequalities delineate the region of the parameter space in which the model is physically acceptable.

For the benchmark in Sect.~\ref{sec:qlaw} and the figure captions, Table~\ref{tab:viability} summarizes, over $-0.9\le z\le3$, which viability conditions are satisfied, based on the computations of Figs.~\ref{fig:rhoG}--\ref{fig:rhom}, \ref{fig:Omega-three-panel-exact}, and \ref{fig:vs2}.

\begin{table}[htbp]
\centering
\footnotesize
\caption{Summary of physical viability tests for the benchmark parameter set, evaluated over $-0.9\le z\le3$. A condition is marked \checkmark\ where it holds for the entire range and $\times$ where it fails over part of the range, with the failing sub-range given explicitly.}
\label{tab:viability}
\begin{tabular}{lll}
\hline\noalign{\smallskip}
Condition & Status & Remark \\
\noalign{\smallskip}\hline\noalign{\smallskip}
$\rho_{G}>0$ & \checkmark & Holds for all $z\in[-0.9,3]$ \\
$\rho_{m}>0$ & $\times$ & Fails for $z\lesssim-0.324$; holds for $z\gtrsim-0.324$ \\
$1+f_{\TT}>0$ & \checkmark & Holds for all $z\in[-0.9,3]$; $1+f_{\TT}\in[2.000,2.009]$ \\
$\Om_{m}+\Om_{G}+\Om_{\TT}=1$ & \checkmark & Satisfied identically (Eq.~\eqref{eq:constraint}) \\
$v_{s}^{2}>0$ & $\times$ & Fails for $z\lesssim0.991$ (Fig.~\ref{fig:vs2}) \\
\noalign{\smallskip}\hline
\end{tabular}
\end{table}

Three conditions are satisfied unconditionally for this benchmark: $\rho_{G}>0$ throughout, $1+f_{\TT}$ remains safely positive ($\in[2.000,\,2.009]$), and the closure relation holds identically. The matter density fails for $z\lesssim-0.324$, placing a constraint on the power-law parameters analogous to those reported for related constructions \cite{mandal_temporal_2020,hatkar_topological_2025}, and $v_{s}^{2}<0$ for $z\lesssim0.991$ as discussed above. Both failing conditions must be resolved by a systematic exploration of $(\alpha,\beta,m,c^{2},\eta)$ parameter space, to be addressed jointly with the observational analysis in Sect.~\ref{sec:obsconstraints}.

\section{Phase-Space Analysis}
\label{sec:phasespace}

To gain insight into the global dynamical behaviour of the model independent of any particular initial condition, it is useful to recast the cosmological equations as an autonomous dynamical system and study its critical points \cite{mirza_constraining_2017,sharif_phase_2015,halder_phase_2024,rana_phase_2026}. We use the number of $e$-folds $N=\ln a$ as the dynamical time variable, so that, from Eq.~\eqref{eq:Hdot-A},
\be
H'\equiv\frac{dH}{dN}=\frac{\dot H}{H}=\nu-AH.
\label{eq:Hprime}
\ee
We introduce the dimensionless variables
\be
h=\frac{H}{H_{0}},
\qquad
\bar\nu=\frac{\nu}{H_{0}},
\ee
in terms of which Eq.~\eqref{eq:Hprime} becomes
\be
h'=\bar\nu-Ah.
\label{eq:hprime}
\ee
For the matter density parameter, defined as $x=\Om_{m}=\kappa^{2}\rho_{m}/(3H^{2})$, standard dust conservation $\rho_{m}'+3\rho_{m}=0$ gives, in terms of the dimensionless variables,
\be
x'=x\Big[-3-2\Big(\frac{\bar\nu}{h}-A\Big)\Big].
\label{eq:xprime}
\ee
The pair of Eqs.~\eqref{eq:hprime} and \eqref{eq:xprime} together form the reduced two-dimensional autonomous system
\be
\begin{cases}
h'=\bar\nu-Ah,\\[2mm]
x'=x\Big[-3-2\Big(\dfrac{\bar\nu}{h}-A\Big)\Big],
\end{cases}
\label{eq:autonomous}
\ee
while the GHRDE density parameter is fixed, at every point of the phase space, by the closure relation
\be
\Om_{G}=1-x-\Om_{\TT}.
\ee

\subsection{Critical Points and Linear Stability}

The critical points of the autonomous system \eqref{eq:autonomous} are obtained by simultaneously imposing $h'=0$ and $x'=0$. From $h'=0$,
\be
h_{c}=\frac{\bar\nu}{A},
\ee
which corresponds, in dimensionful terms, to $H_{c}=\nu/A$ and to a deceleration parameter
\be
q_{c}=b-\frac{\nu}{H_{c}}=b-A=-1,
\ee
confirming that this critical point describes a de~Sitter accelerated phase, consistent with the late-time limit identified in Sect.~\ref{sec:qlaw}.

From $x'=0$, either
\be
x_{c}=0,
\ee
corresponding to a fully dark-energy-dominated Universe, or
\be
-3-2\Big(\frac{\bar\nu}{h_{c}}-A\Big)=0
\quad\Longrightarrow\quad
\frac{\bar\nu}{h_{c}}=A-\frac32.
\ee
Using $q=b-\bar\nu/h$, the second branch corresponds to
\be
q=b-\Big(A-\frac32\Big)=\frac12,
\ee
which is the characteristic value of the deceleration parameter for a matter-dominated decelerating Universe. Note that the condition defining $P_{m}$ fixes $h$ alone, $h_{m}=\bar\nu/(A-3/2)$, while leaving $x$ unconstrained; $P_{m}$ is therefore not an isolated point but a critical \emph{line} $\{(h_{m},x):x\ge0\}$ of the autonomous system.

To determine the stability of each critical object, we linearize the autonomous system~\eqref{eq:autonomous} and evaluate the Jacobian matrix
\be
J=\begin{pmatrix}
\dfrac{\partial h'}{\partial h} & \dfrac{\partial h'}{\partial x}\\[2mm]
\dfrac{\partial x'}{\partial h} & \dfrac{\partial x'}{\partial x}
\end{pmatrix}
=\begin{pmatrix}
-A & 0\\[2mm]
\dfrac{2\bar\nu x}{h^{2}} & -3-2\Big(\dfrac{\bar\nu}{h}-A\Big)
\end{pmatrix}.
\label{eq:jacobian}
\ee
Since $h'$ does not depend on $x$, $J$ is lower-triangular at every point of the phase space, so its eigenvalues are read off directly from the diagonal.

At $P_{dS}$ ($h=h_{c}=\bar\nu/A$, $x=0$), the diagonal entries evaluate to
\be
\lambda_{1}=-A,\qquad \lambda_{2}=-3-2\Big(\frac{\bar\nu}{h_{c}}-A\Big)=-3.
\ee
Both eigenvalues are negative for any $A>0$, so $P_{dS}$ is a \textbf{stable node} (attractor), confirming that the de~Sitter solution is approached asymptotically regardless of the initial conditions in $h$ and $x$.

At $P_{m}$ ($h=h_{m}=\bar\nu/(A-3/2)$, $x$ arbitrary), the diagonal entries evaluate to
\be
\lambda_{1}=-A,\qquad \lambda_{2}=-3-2\Big(\frac{\bar\nu}{h_{m}}-A\Big)=0,
\ee
where $\lambda_{2}=0$ is identical to the defining condition of $P_{m}$. The eigenvalue pair $(-A,0)$ does \emph{not} correspond to a saddle; rather, $P_{m}$ is a \textbf{non-hyperbolic} critical object, stable in the $h$-direction but marginal in the $x$-direction, consistent with $P_{m}$ being a line of fixed points rather than an isolated point. Table~\ref{tab:critical} summarizes both critical objects, along with their eigenvalues and stability classification.

\begin{table}[htbp]
\centering
\footnotesize
\caption{Critical points of the autonomous system~\eqref{eq:autonomous}, their eigenvalues, and cosmological interpretation.}
\label{tab:critical}
\begin{tabular}{ccccc}
\hline\noalign{\smallskip}
Point & Condition & $q$ & Eigenvalues & Classification\\
\noalign{\smallskip}\hline\noalign{\smallskip}
$P_{m}$ & $q=1/2$ & $1/2$ & $(-A,\,0)$ & Non-hyperbolic ($q=1/2$ decelerating branch)\\
$P_{dS}$ & $H=\nu/A$ & $-1$ & $(-A,\,-3)$ & Stable node (de~Sitter attractor)\\
\noalign{\smallskip}\hline
\end{tabular}
\end{table}

\begin{figure}[htbp]
\centering
\includegraphics[width=\linewidth]{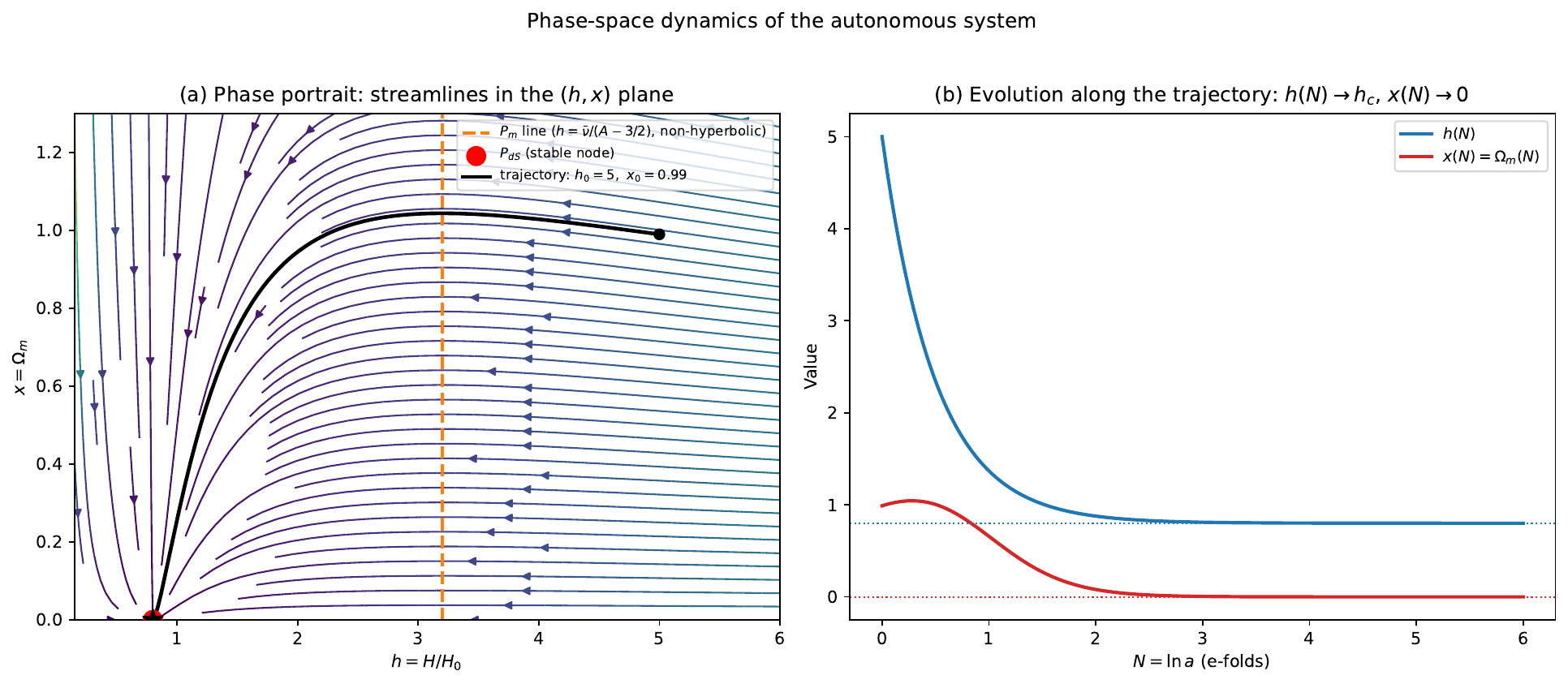}
\caption{Phase-space dynamics of the autonomous system~\eqref{eq:autonomous} for $A=2$, $\bar\nu=1.6$ (i.e.\ $H_{0}=69$ km s$^{-1}$ Mpc$^{-1}$, $\nu=110.4$ km s$^{-1}$ Mpc$^{-1}$). (a) Streamlines in the $(h,x)$ plane, with the non-hyperbolic critical line $P_{m}$ ($h=h_{m}=3.2$, dashed) and the stable node $P_{dS}$ ($h=h_{c}=0.8$, $x=0$, red dot) marked, together with a representative trajectory launched from $(h_{0},x_{0})=(5,0.99)$. (b) The same trajectory shown as $h(N)$ and $x(N)=\Om_{m}(N)$ versus the number of $e$-folds $N=\ln a$, converging to $h_{c}=0.8$ and $x=0$.}
\label{fig:phaseportrait}
\end{figure}

Figure~\ref{fig:phaseportrait} shows the resulting phase portrait together with a representative trajectory launched from $(h_{0},x_{0})=(5,0.99)$, a strongly matter-dominated, high-$H$ initial condition. As panel (a) shows, this trajectory passes close to the $P_{m}$ line (it crosses $h=h_{m}$ at $N\approx0.3$, where it reaches its maximum value $x\approx1.04$) before being driven monotonically towards $P_{dS}$, which it approaches to within $|x|\sim10^{-6}$ by $N\approx5$--$6$, exactly as expected from a stable node with eigenvalues $(-A,-3)$. Because $x$ here represents the effective matter density parameter of the reduced dynamical system rather than an observationally normalized density fraction, values slightly above unity along the trajectory do not indicate a physical inconsistency; they simply reflect the transient redistribution between the matter and GHRDE sectors required by the closure relation at that point in the phase space. Because $P_{m}$ is non-hyperbolic rather than a saddle, the trajectory is not deflected away from it along a genuine unstable manifold; instead, the $x$-direction is only marginally stable at $P_{m}$ to linear order, so the trajectory's residence near $h_{m}$ is governed by the (here mild) nonlinear terms in Eq.~\eqref{eq:xprime}, rather than by exponential repulsion. Panel (b) makes this concrete: $x(N)$ rises slightly and then decays smoothly to zero as $h(N)$ relaxes to $h_{c}$, with no abrupt or discontinuous behaviour near $N\approx0.3$.

The phase portrait reproduces the cosmological sequence required of a viable dark-energy model: trajectories beginning with large $\Omega_m$ transit through the neighbourhood of $P_{m}$ ($q=1/2$) before converging to $P_{dS}$ ($q=-1$, de~Sitter). Both eigenvalues of $P_{dS}$ being strictly negative for any $A>0$ guarantees this convergence for generic initial conditions, which is a stronger viability statement than that for a saddle or unstable-node fixed point. The same attractor structure has been identified in the dynamical analyses of related $f(\TT)$ and modified-gravity dark-energy models \cite{mirza_constraining_2017,sharif_phase_2015,halder_phase_2024,rana_phase_2026}.

\section{Energy Conditions}
\label{sec:energyconditions}

Energy conditions provide model-independent constraints on the effective energy density and pressure of a cosmological fluid and are widely used as consistency checks for modified gravity scenarios \cite{sharif_dark_2014,zubair_thermodynamic_2015,setare_cosmological_2012}. Since the dust component is pressureless, $p_{m}=0$, the total effective density and pressure are $\rho_{\rm eff}=\rho_{m}+\rho_{G}$ and $p_{\rm eff}=p_{G}$, with $\rho_{m}$ and $p_{G}$ given by Eqs.~\eqref{eq:rhom-final} and \eqref{eq:pG-final} respectively. The null, weak, dominant, and strong energy conditions (NEC, WEC, DEC, SEC) are defined, respectively, by
\be
\begin{aligned}
&\text{NEC:}\quad \rho_{\rm eff}+p_{\rm eff}\geq0,\\
&\text{WEC:}\quad \rho_{\rm eff}\geq0 \;\&\; \rho_{\rm eff}+p_{\rm eff}\geq0,\\
&\text{DEC:}\quad \rho_{\rm eff}\geq0 \;\&\; \rho_{\rm eff}\pm p_{\rm eff}\geq0,\\
&\text{SEC:}\quad \rho_{\rm eff}+p_{\rm eff}\geq0 \;\&\; \rho_{\rm eff}+3p_{\rm eff}\geq0.
\end{aligned}
\label{eq:ec-defs}
\ee
Substituting Eqs.~\eqref{eq:fT}, \eqref{eq:fTT}, \eqref{eq:Tscalar}, and \eqref{eq:Hdot-A} into the expressions for $\rho_{\rm eff}$ and $p_{\rm eff}$, the four condition-defining combinations reduce, after simplification, to
\be
\begin{aligned}
\rho_{\rm eff}+p_{\rm eff}
=\frac{H(\nu-AH)}{2\kappa^{2}}\Big[&
48H^{2}\beta m(m-1)\TT^{m-2}\\
&-4\big(1+\alpha+\beta m\TT^{m-1}\big)
\Big],
\end{aligned}
\label{eq:nec-H}
\ee
\be
\begin{aligned}
\rho_{\rm eff}=\frac{1}{2\kappa^{2}}\Big[&6H^{2}+12H^{2}\big(\alpha+\beta m\TT^{m-1}\big)\\
&+\alpha\TT+\beta\TT^{m}\Big],
\end{aligned}
\label{eq:wec-rho-model}
\ee
\be
\begin{aligned}
\rho_{\rm eff}-p_{\rm eff}
=\frac{1}{2\kappa^{2}}\Big[&
-48H^{3}(\nu-AH)\beta m(m-1)\TT^{m-2}\\
&+4H(\nu-AH)+12H^{2}\\
&+\big\{24H^{2}+4H(\nu-AH)\big\}\\
&\quad\times\big(\alpha+\beta m\TT^{m-1}\big)\\
&+2\big(\alpha\TT+\beta\TT^{m}\big)
\Big],
\end{aligned}
\label{eq:dec-H}
\ee
\be
\begin{aligned}
\rho_{\rm eff}+3p_{\rm eff}
=\frac{1}{2\kappa^{2}}\Big[&
144H^{3}(\nu-AH)\beta m(m-1)\TT^{m-2}\\
&-12H(\nu-AH)-12H^{2}\\
&-\big\{24H^{2}+12H(\nu-AH)\big\}\\
&\quad\times\big(\alpha+\beta m\TT^{m-1}\big)\\
&-2\big(\alpha\TT+\beta\TT^{m}\big)
\Big],
\end{aligned}
\label{eq:sec-H}
\ee
Each of these expressions depends on $H$, and hence on the redshift through Eq.~\eqref{eq:Hz}, so its sign can in principle change over cosmic history. For an accelerating Universe, a violation of the SEC is expected and is not, by itself, a pathology, since it is precisely the violation of the SEC that drives cosmic acceleration in both $\Lambda$CDM and in most modified-gravity dark-energy models. In contrast, a violation of the NEC must be examined carefully, as it may signal phantom-like behaviour or a possible classical or quantum instability.

\begin{figure}[htbp]
\centering
\includegraphics[width=\linewidth]{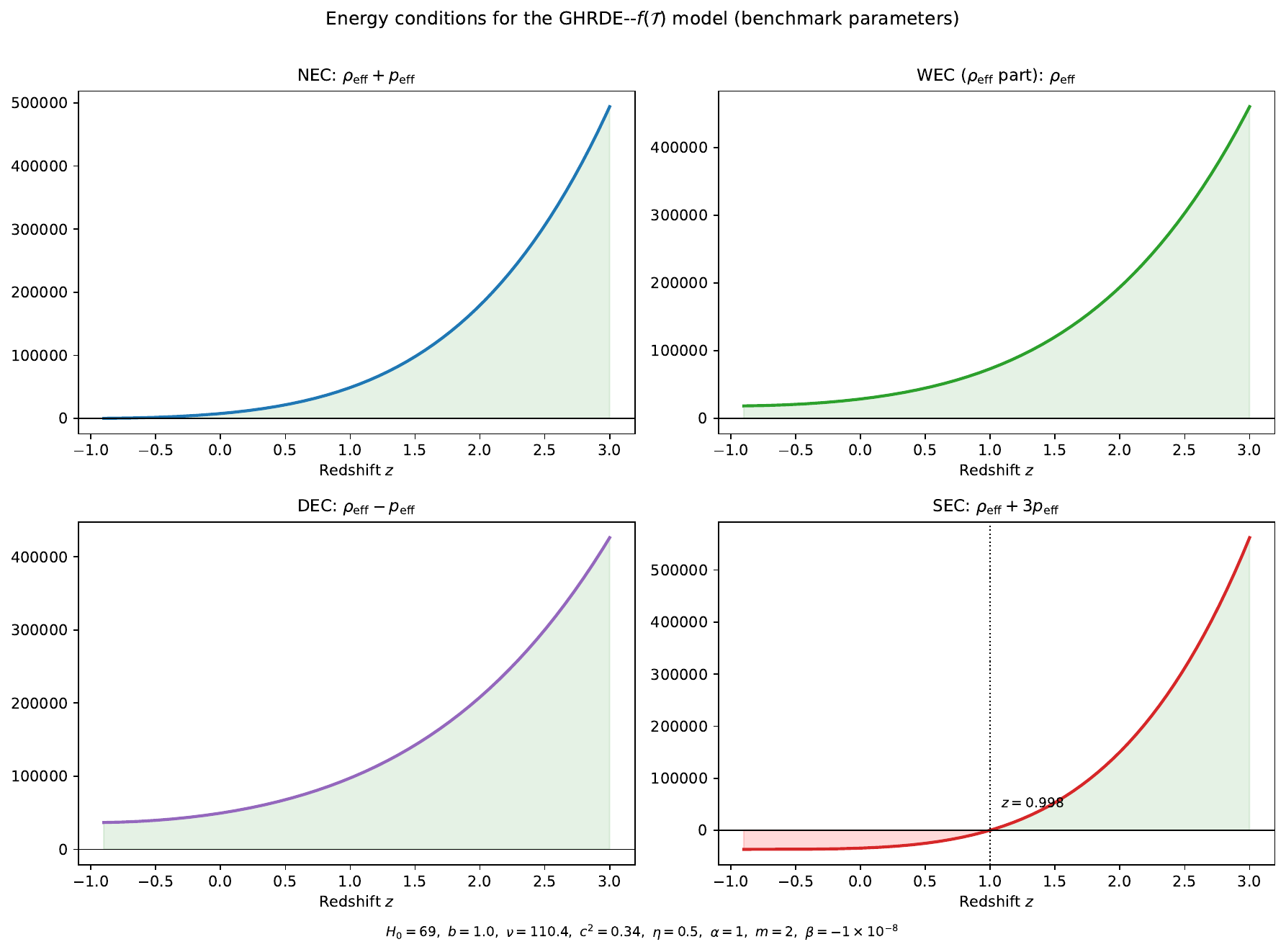}
\caption{The four energy-condition combinations $E_{\rm NEC}=\rho_{\rm eff}+p_{\rm eff}$ (Eq.~\eqref{eq:nec-H}), $\rho_{\rm eff}$ (Eq.~\eqref{eq:wec-rho-model}, the additional WEC requirement beyond the NEC), $E_{\rm DEC}^{(-)}=\rho_{\rm eff}-p_{\rm eff}$ (Eq.~\eqref{eq:dec-H}), and $E_{\rm SEC}=\rho_{\rm eff}+3p_{\rm eff}$ (Eq.~\eqref{eq:sec-H}), evaluated over $-0.9\le z\le3$ for the benchmark parameter set: $H_{0}=69$ km s$^{-1}$ Mpc$^{-1}$, $b=1.0$, $\nu=110.4$ km s$^{-1}$ Mpc$^{-1}$, $c^{2}=0.34$, $\eta=0.5$, $\alpha=1$, $m=2$, $\beta=-1\times10^{-8}$. Green/red shading marks the regions where each quantity is non-negative/negative; the dotted vertical line in the SEC panel marks the sign-change redshift $z=0.998$.}
\label{fig:energy}
\end{figure}

Figure~\ref{fig:energy} shows the four energy-condition combinations for the benchmark parameter set. The NEC combination $\rho_{\rm eff}+p_{\rm eff}$ (top-left panel) is strictly positive throughout $-0.9\le z\le3$, confirming the absence of phantom-like behaviour. The effective energy density $\rho_{\rm eff}$ (top-right panel) is likewise positive and grows monotonically with redshift, consistent with the expected increase of the total energy budget in the past; this simultaneously confirms the $\rho_{\rm eff}\ge0$ requirement of the WEC beyond what the NEC alone provides. The DEC combination $\rho_{\rm eff}-p_{\rm eff}$ (bottom-left panel) remains positive throughout, increasing steeply toward higher redshift as $\rho_{\rm eff}$ grows faster than $|p_{\rm eff}|$; the DEC is therefore satisfied unconditionally. In contrast, the SEC combination $\rho_{\rm eff}+3p_{\rm eff}$ (bottom-right panel) is negative for $z\lesssim0.998$ and turns positive only beyond this redshift: the red-shaded region covers the entire present and near-future epoch, while the green-shaded region begins near $z=0.998$, coinciding closely with the deceleration-to-acceleration transition $z_{t}=1.0$ identified from $q(z)$ in Sect.~\ref{sec:qlaw}.

For this benchmark, NEC, WEC, and DEC are satisfied throughout the entire range $-0.9\le z\le3$. The SEC is violated for $z\lesssim0.998$, which is the standard general-relativistic signature of accelerated expansion (via Raychaudhuri's equation) rather than an independent pathology, and is entirely consistent with the independent evidence from $q_{0}<0$ and $\omega_{\rm eff,0}<-1/3$.

The NEC satisfaction throughout does not contradict the negative $v_{s}^{2}$ for $z\lesssim0.991$ or the negative $\rho_{m}$ for $z\lesssim-0.324$ identified in Sect.~\ref{sec:stability}: energy conditions test the total effective fluid $\rho_{\rm eff}=\rho_{m}+\rho_{G}$, for which the dominant GHRDE contribution keeps the sum non-negative \cite{sharif_dark_2014,zubair_thermodynamic_2015}, whereas the stability and viability conditions in Table~\ref{tab:viability} test individual sectors.

\section{Observational Methodology and Future Constraints}
\label{sec:obsconstraints}

The model developed above can, in principle, be confronted with current cosmological observations through the redshift-space Hubble function
\be
H(z)=\frac{\nu}{A}+\Big(H_{0}-\frac{\nu}{A}\Big)(1+z)^{A},
\qquad A=1+b,
\ee
which depends on the full parameter vector
\be
\bm{\vartheta}=\{H_{0},\,b,\,\nu,\,\alpha,\,\beta,\,m,\,c^{2},\,\eta\}.
\ee
Since several of these parameters are strongly correlated --- in particular $\alpha$ with $\beta$ and $m$ --- a first observational analysis may fix $\alpha=1$ (GR-normalized value) and constrain only the reduced set
\be
\bm{\vartheta}_{\rm red}=\{H_{0},\,b,\,\nu,\,\beta,\,m,\,c^{2},\,\eta\}.
\ee
We outline below the three main observational probes --- cosmic chronometers, Pantheon$+$ Type~Ia supernovae, and BAO --- through which $\bm{\vartheta}_{\rm red}$ can be constrained.

\subsection{Cosmic Chronometer \texorpdfstring{$H(z)$}{H(z)} Data}

Cosmic chronometers yield direct, model-independent $H(z)$ measurements from the relative ages of passively evolving galaxies \cite{jimenez_constraints_2002,moresco_raising_2016,nunes_new_2017,sudharani_probing_2024}. Given $N_{H}$ measurements $H_{\rm obs}(z_{i})$ with uncertainties $\sigma_{H,i}$, the chi-square is
\be
\chi^{2}_{H}=\sum_{i=1}^{N_{H}}
\frac{\big[H_{\rm th}(z_{i})-H_{\rm obs}(z_{i})\big]^{2}}{\sigma_{H,i}^{2}},
\ee
with $H_{\rm th}(z_{i})$ being the theoretical Hubble parameter of Eq.~\eqref{eq:Hz}.

\subsection{Pantheon+ Supernova Data}

Type Ia supernovae constrain the apparent distance modulus $\mu(z)$, which is computed from the luminosity distance $D_{L}(z)=c(1+z)\int_{0}^{z}dz'/H(z')$. Denoting by $\Delta\bm{\mu}$ the residuals over the full Pantheon$+$ sample \cite{brout_pantheon_2022,scolnic_pantheon_2022} and by $C$ the covariance matrix, the supernova chi-square is
\be
\chi^{2}_{SN}=\Delta\bm{\mu}^{T}C^{-1}\Delta\bm{\mu}.
\ee

\subsection{BAO Data}

Baryon acoustic oscillation surveys \cite{beutler_6df_2011,alam_boss_2017} constrain combinations of the comoving angular diameter distance $D_{M}(z)$, Hubble distance $D_{H}(z)=c/H(z)$, and spherically averaged distance $D_{V}(z)=[zD_{M}^{2}(z)D_{H}(z)]^{1/3}$. Denoting by $\bm{X}$ the vector of BAO observables and $C_{BAO}$ the covariance matrix, the BAO chi-square is
\be
\chi^{2}_{BAO}=\Delta\bm{X}^{T}C^{-1}_{BAO}\Delta\bm{X}.
\ee

\subsection{Total Likelihood}

Assuming the three datasets are statistically independent, the total chi-square is the simple sum
\be
\chi^{2}_{\rm total}=\chi^{2}_{H}+\chi^{2}_{SN}+\chi^{2}_{BAO},
\ee
with the corresponding likelihood $\mathcal{L}\propto\exp(-\chi^{2}_{\rm total}/2)$. Minimizing $\chi^{2}_{\rm total}$ over $\bm{\vartheta}_{\rm red}$ yields the best-fit values and credible intervals of $H_{0}$, $b$, $\nu$, $\beta$, $m$, $c^{2}$, and $\eta$.

In this paper, we present the theoretical framework, closed-form redshift-space expressions, and qualitative dynamical and stability diagnostics of the GHRDE--$f(\TT)$ model. A full MCMC analysis combining CC, Pantheon$+$ \cite{brout_pantheon_2022,scolnic_pantheon_2022}, and BAO \cite{beutler_6df_2011,alam_boss_2017} datasets, yielding the best-fit values and credible intervals for $\bm{\vartheta}_{\rm red}$, will be the subject of a dedicated future study.

\section{Conclusions}
\label{sec:conclusions}

This paper has presented a self-contained analytical study of a cosmological model in
which generalized holographic Ricci dark energy of Xu type is embedded within
$f(\TT)=\alpha\TT+\beta\TT^{m}$ teleparallel gravity.  The primary novelty of the
construction lies in driving the late-time dynamics entirely through a
Hubble-parameter-dependent deceleration law, $q=b-\nu/H$, rather than by specifying
the scale factor a priori.  This single parametric choice propagates analytically
through every level of the analysis, yielding a closed-form redshift-space Hubble
function $H(z)=\nu/A+(H_{0}-\nu/A)(1+z)^{A}$ and enabling the derivation of the GHRDE
energy density, pressure, and all subsequent cosmological diagnostics as explicit
functions of $H$, and hence of the redshift.

The principal results, obtained with the illustrative benchmark
$H_{0}=69$ km\,s$^{-1}$\,Mpc$^{-1}$, $b=1.0$, $\nu=110.4$ km\,s$^{-1}$\,Mpc$^{-1}$,
$c^{2}=0.34$, $\eta=0.5$, $\alpha=1$, $m=2$, $\beta=-10^{-8}$, are summarised below.

\textit{Expansion history.}  The Hubble parameter decreases monotonically from
the past toward the future and asymptotes to $H_{\infty}=55.2$ km\,s$^{-1}$\,Mpc$^{-1}$
as $z\to-1$, corresponding to an exact de~Sitter phase at late times.  The deceleration
parameter crosses zero at the transition redshift $z_{t}=1.0$, with the present-day
value $q_{0}=-0.60$ lying within the range $-0.55$ to $-0.71$ reported for
comparable modified teleparallel gravity analyses
\cite{pal_cosmological_2025,bhoyar_resolving_2024,goswami_modeling_2021}.

\textit{GHRDE fluid and cosmological parameters.} The GHRDE equation-of-state parameter $\omega_{G,0}=-0.81$ places the dark-energy sector in the quintessence band $-1<\omega_{G}<-1/3$. The effective equation of state $\omega_{\rm eff,0}=-0.733$ lies below the acceleration threshold $-1/3$ and evolves toward $-1$ in the asymptotic future without crossing the phantom divide. The three density parameters satisfy $\Omega_{m}+\Omega_{G}+\Omega_{\TT}=1$ identically (verified to $2\times10^{-16}$).

\textit{Statefinder diagnostic.}  The geometrical statefinder pair
\cite{sahni_statefinder_2003,alam_exploring_2003} evaluated at the present epoch gives
$(r_{0},s_{0})=(0.760,\,0.073)$, situating the model in the quintessence region
($r<1$, $s>0$) of the $r$--$s$ plane.  The trajectory passes exactly through the
$\Lambda$CDM fixed point $(r,s)=(1,0)$ at $z=1$, where $q=0$, and a structurally
inevitable pole appears at $z_{p}=2.464$ where $q=1/2$, arising from the denominator
of the standard $s$-definition rather than from any pathology of the underlying
cosmology \cite{enkhili_diagnostic_2024}.

\textit{Classical stability.}  The squared adiabatic sound speed is
$v_{s}^{2}(0)=-0.406$, indicating a classical instability of linear perturbations in
the GHRDE fluid at and below the present epoch; $v_{s}^{2}$ changes sign at
$z_{c}=0.991$, becoming positive for $z\gtrsim0.991$.  This sign change originates
in the pressure response of the GHRDE fluid and is a feature shared by several
power-law and exponential $f(\TT)$ constructions in the literature
\cite{bhoyar_stability_2017,paliathanasis_stability_2018}.  The viability assessment
further shows that while $\rho_{G}>0$ and $1+f_{\TT}\in[2.000,\,2.009]$ are satisfied
throughout, the matter density turns negative in the far-future regime
$z\lesssim-0.324$, placing a genuine constraint on the combination of power-law
exponent and deviation parameter.

\textit{Phase-space dynamics.}  The reduced autonomous system in the $(h,x)$ plane
possesses two critical objects: a non-hyperbolic critical line $P_{m}$ at $q=1/2$,
characterising a matter-dominated decelerating configuration, and a stable-node attractor
$P_{dS}$ at $q=-1$ with eigenvalues $(-A,-3)$, both strictly negative for any $A>0$.
Generic trajectories originating in matter-dominated initial conditions transit through
the neighbourhood of $P_{m}$ before converging irreversibly to $P_{dS}$, reproducing
the late-time cosmological sequence required of a viable dark-energy model and
consistent with dynamical analyses of related $f(\TT)$ constructions
\cite{mirza_constraining_2017,sharif_phase_2015,halder_phase_2024,rana_phase_2026}.

\textit{Energy conditions.}  The null, weak, and dominant energy conditions are satisfied
throughout the redshift range $-0.9\le z\le3$ for this benchmark.  The strong energy
condition is violated for $z\lesssim0.998$, coinciding with the onset of cosmic
acceleration at $z_{t}=1.0$; this violation is the expected general-relativistic
signature of accelerated expansion and is entirely consistent with the independent
evidence from $q_{0}<0$ and $\omega_{\rm eff,0}<-1/3$.

Taken together, these results confirm that the GHRDE--$f(\TT)$ framework is a theoretically consistent and analytically tractable platform for late-time dark-energy cosmology. A full MCMC analysis combining cosmic chronometer data \cite{jimenez_constraints_2002,moresco_raising_2016}, Pantheon$+$ supernovae \cite{brout_pantheon_2022,scolnic_pantheon_2022}, and BAO measurements \cite{beutler_6df_2011,alam_boss_2017} will constrain the parameters to a physically viable region, and is the natural continuation of this work.

\section*{Funding}
This research received no specific grant from any funding agency in the public, commercial, or not-for-profit sectors.

\section*{Data Availability}
\noindent Data sharing is not applicable to this article as no datasets were generated or analysed during the current study.


\begin{thebibliography}{99}

\bibitem{riess_observational_1998}
A.~G.~Riess et al., \textit{Observational evidence from supernovae for an accelerating universe and a cosmological constant}, Astron. J. \textbf{116}, 1009--1038 (1998).

\bibitem{perlmutter_measurements_1999}
S.~Perlmutter et al., \textit{Measurements of \ensuremath\Omega and \ensuremath\Lambda from 42 high-redshift supernovae}, Astrophys. J. \textbf{517}, 565--586 (1999).

\bibitem{aghanim_planck_2020}
N.~Aghanim, others, \textit{Planck 2018 results. VI. Cosmological parameters}, Astronomy \& Astrophysics \textbf{641}, A6 (2020).

\bibitem{hayashi_new_1979}
K.~Hayashi, T.~Shirafuji, \textit{New general relativity}, Physical Review D \textbf{19}, 3524--3553 (1979).

\bibitem{maluf_teleparallel_2013}
J.~W.~Maluf, \textit{The teleparallel equivalent of general relativity}, Annalen der Physik \textbf{525}, 339--357 (2013).

\bibitem{ferraro_modified_2007}
R.~Ferraro, F.~Fiorini, \textit{Modified teleparallel gravity: Inflation without inflaton}, Physical Review D \textbf{75}, 084031 (2007).

\bibitem{bengochea_dark_2009}
G.~Bengochea, R.~Ferraro, \textit{Dark torsion as the cosmic speed-up}, Phys. Rev. D \textbf{79}, 124019 (2009).

\bibitem{linder_einsteins_2010}
E.~V.~Linder, \textit{Einstein's Other Gravity and the Acceleration of the Universe}, Phys. Rev. D \textbf{81}, 127301 (2010).

\bibitem{li_ft_2011}
B.~Li, T.~P.~Sotiriou, J.~D.~Barrow, \textit{f(T) Gravity and local Lorentz invariance}, Phys. Rev. D \textbf{83}, 064035 (2011).

\bibitem{cai_ft_2016}
Y.~F.~Cai, S.~Capozziello, M.~D.~Laurentis, E.~N.~Saridakis, \textit{f(T) teleparallel gravity and cosmology}, Rep. Prog. Phys. \textbf{79}, 106901 (2016).

\bibitem{yang_new_2011}
R.~J.~Yang, \textit{New types of $f(T)$ gravity}, Eur. Phys. J. C \textbf{71}, 1797 (2011).

\bibitem{myrzakulov_accelerating_2011}
R.~Myrzakulov, \textit{Accelerating universe from F(T) gravity}, Eur. Phys. J. C \textbf{71}, 1752 (2011).

\bibitem{sharif_ft_2011}
M.~Sharif, S.~Rani, \textit{F(T) Models within Bianchi Type I Universe}, Mod. Phys. Lett. A \textbf{26}, 1657--1671 (2011).

\bibitem{setare_power-law_2012}
M.~R.~Setare, F.~Darabi, \textit{Power-law solutions in $f(T)$ gravity}, Gen Relativ Gravit \textbf{44}, 2521--2527 (2012).

\bibitem{daouda_reconstruction_2012}
M.~H.~Daouda, M.~E.~Rodrigues, M.~J.~S.~Houndjo, \textit{Reconstruction of $f(T)$ gravity according to holographic dark energy}, Eur. Phys. J. C \textbf{72}, 1893 (2012).

\bibitem{rodrigues_locally_2014}
M.~E.~Rodrigues, I.~G.~Salako, M.~J.~S.~Houndjo, J.~Tossa, \textit{Locally Rotationally Symmetric Bianchi Type-I cosmological model in $f(T)$ gravity: from early to Dark Energy dominated universe}, Int. J. Mod. Phys. D \textbf{23}, 1450004 (2014).

\bibitem{capozziello_model-independent_2017}
S.~Capozziello, R.~D’Agostino, O.~Luongo, \textit{Model-independent reconstruction of $f(T)$ teleparallel cosmology}, Gen Relativ Gravit \textbf{49}, 141 (2017).

\bibitem{sharif_thermodynamics_2014}
M.~Sharif, S.~Rani, \textit{Thermodynamics in $f(T)$ Gravity and Corrected Entropies}, arXiv:1412.5116 [gr-qc] (2014).

\bibitem{karami_generalized_2012}
K.~Karami, A.~Abdolmaleki, \textit{Generalized second law of thermodynamics in $f(T)$ gravity}, J. Cosmol. Astropart. Phys. \textbf{2012}, 007--007 (2012).

\bibitem{mirza_constraining_2017}
B.~Mirza, F.~Oboudiat, \textit{Constraining $f(T)$ gravity by dynamical system analysis}, J. Cosmol. Astropart. Phys. \textbf{2017}, 011--011 (2017).

\bibitem{sharif_phase_2015}
M.~Sharif, S.~Jabbar, \textit{Phase Space Analysis and Anisotropic Universe Model in $f(T)$ Gravity}, Commun. Theor. Phys. \textbf{63}, 168--180 (2015).

\bibitem{paliathanasis_stability_2018}
A.~Paliathanasis, J.~L.~Said, J.~D.~Barrow, \textit{Stability of the Kasner universe in $f(T)$ gravity}, Phys. Rev. D \textbf{97}, 044008 (2018).

\bibitem{halder_phase_2024}
S.~Halder, \textit{Phase space analysis of sign-shifting interacting dark energy models}, Phys. Rev. D \textbf{109} (2024).

\bibitem{rana_phase_2026}
D.~S.~Rana, R.~Solanki, P.~K.~Sahoo, \textit{Phase space analysis of the viscous fluid cosmological models in the coincident $f(T)$ gravity}, Annals of Physics \textbf{492}, 170567 (2026).

\bibitem{wu_observational_2010}
P.~Wu, H.~Yu, \textit{Observational constraints on f (T) theory}, Physics Letters B \textbf{693}, 415--420 (2010).

\bibitem{nunes_new_2018}
R.~C.~Nunes, S.~Pan, E.~N.~Saridakis, \textit{New observational constraints on $f(T)$ gravity through gravitational-wave astronomy}, Phys. Rev. D \textbf{98}, 104055 (2018).

\bibitem{santos_observational_2022}
F.~B.~M.~d.~Santos, J.~E.~Gonzalez, R.~Silva, \textit{Observational Constraints on $f(T)$ Gravity from Model-Independent Data}, Eur. Phys. J. C \textbf{82}, 823 (2022).

\bibitem{capozziello_transition_2015}
S.~Capozziello, O.~Luongo, E.~N.~Saridakis, \textit{Transition redshift in $f(T)$ cosmology and observational constraints}, Phys. Rev. D \textbf{91}, 124037 (2015).

\bibitem{escamilla-rivera_ft_2024}
C.~Escamilla-Rivera, R.~Sandoval-Orozco, \textit{$f(T)$ gravity after DESI Baryon Acoustic Oscillation and DES Supernovae 2024 data}, Journal of High Energy Astrophysics \textbf{42}, 217--221 (2024).

\bibitem{jiang_exploring_2024}
X.~Jiang et al., \textit{Exploring $f(T)$ Gravity via strongly lensed fast radio bursts}, arXiv:2401.05464 [gr-qc] (2024).

\bibitem{saridakis_introduction_2017}
E.~N.~Saridakis, \textit{Introduction to teleparallel and $f(T)$ gravity and cosmology}, preprint, doi:10.1142/9789813226609\_0074 (2017).

\bibitem{bahamonde_teleparallel_2023}
S.~Bahamonde et al., \textit{Teleparallel Gravity: From Theory to Cosmology}, Rep. Prog. Phys. \textbf{86}, 026901 (2023).

\bibitem{li_model_2004}
M.~Li, \textit{A Model of Holographic Dark Energy}, Physics Letters B \textbf{603}, 1--5 (2004).

\bibitem{huang_holographic_2004}
Q.~G.~Huang, M.~Li, \textit{The Holographic Dark Energy in a Non-flat Universe}, J. Cosmol. Astropart. Phys. \textbf{2004}, 013--013 (2004).

\bibitem{gao_holographic_2009}
C.~Gao, F.~Wu, X.~Chen, Y.-G.~Shen, \textit{Holographic dark energy model from Ricci scalar curvature}, Physical Review D \textbf{79}, 043511 (2009).

\bibitem{granda_infrared_2008}
L.~N.~Granda, A.~Oliveros, \textit{Infrared cut-off proposal for the holographic density}, Phys. Lett. B \textbf{671}, 199--202 (2009).

\bibitem{xu_generalized_2009}
L.~Xu, J.~Lu, W.~Li, \textit{Generalized Holographic and Ricci Dark Energy Models}, Eur. Phys. J. C \textbf{64}, 89 (2009).

\bibitem{lu_cosmological_2012}
J.~Lu, Y.~Wang, Y.~Wu, T.~Wang, \textit{Cosmological constraints on the generalized holographic dark energy}, Phys. Lett. B \textbf{713}, 6--13 (2012).

\bibitem{bhattacharya_study_2012}
S.~Bhattacharya, U.~Debnath, \textit{Study of Thermodynamics in Generalized Holographic and Ricci Dark Energy Models}, Int J Theor Phys \textbf{51}, 577--588 (2012).

\bibitem{enkhili_diagnostic_2024}
O.~Enkhili et al., \textit{Diagnostic Approaches for Interacting generalized holographic Ricci Dark Energy Models}, New Astronomy \textbf{113}, 102298 (2024).

\bibitem{pasqua_generalized_2025}
A.~Pasqua, \textit{Generalized Holographic and Ricci Dark Energy: Cosmological Diagnostics and Scalar Field Realizations}, arXiv:2509.19386 [gr-qc] (2025).

\bibitem{sharif_comparative_2026}
M.~Sharif, M.~Z.~Gul, I.~Hashim, \textit{Comparative Analysis of Holographic Dark Energy Models in $f(R,T^2)$ Gravity}, arXiv:2602.03898 [gr-qc] (2026).

\bibitem{chirde_dynamic_2018}
V.~R.~Chirde, S.~H.~Shekh, \textit{Dynamic minimally interacting holographic dark energy cosmological model in $f(T)$ gravity}, Indian J Phys \textbf{92}, 1485--1494 (2018).

\bibitem{bhardwaj_renyi_2022}
V.~K.~Bhardwaj, A.~Dixit, A.~Pradhan, S.~Krishannair, \textit{Renyi Holographic Dark Energy models in Teleparallel gravity}, Int. J. Mod. Phys. A \textbf{37}, 2250178 (2022).

\bibitem{koussour_bianchi_2022}
M.~Koussour, S.~H.~Shekh, M.~Bennai, \textit{Bianchi type-I Barrow holographic dark energy model in symmetric teleparallel gravity}, Int. J. Mod. Phys. A \textbf{37}, 2250184 (2022).

\bibitem{dhore_study_2024}
A.~O.~Dhore, M.~R.~Ugale, \textit{Study of $f(T)$ Theory of Gravity in the Framework of Modified Holographic Ricci Dark Energy with Thermodynamical Aspects}, Int J Theor Phys \textbf{63}, 202 (2024).

\bibitem{hatkar_topological_2025}
S.~P.~Hatkar, D.~P.~Tadas, S.~D.~Katore, \textit{Topological defects in $f(T)$ theory of gravity}, Eur. Phys. J. C \textbf{85}, 150 (2025).

\bibitem{pal_cosmological_2025}
S.~Pal, R.~Garg, G.~P.~Singh, G.~Shanker, \textit{Cosmological Dynamics of Accelerating Model in $f(T)$ Gravity with Special Forms of Deceleration Parameter}, arXiv:2506.03756 [gr-qc] (2025).

\bibitem{bhoyar_resolving_2024}
S.~R.~Bhoyar, Y.~B.~Ingole, \textit{Resolving FLRW cosmology through effective equations of state in $f(T)$ gravity}, Chinese Journal of Physics \textbf{92}, 1085--1096 (2024).

\bibitem{bhoyar_stability_2017}
S.~R.~Bhoyar, V.~R.~Chirde, S.~H.~Shekh, \textit{Stability of Accelerating Universe with Linear Equation of State in $f(T)$ Gravity Using Hybrid Expansion Law}, Astrophysics \textbf{60}, 259--272 (2017).

\bibitem{goswami_modeling_2021}
G.~K.~Goswami, A.~K.~Yadav, B.~Mishra, S.~K.~Tripathy, \textit{Modeling of accelerating Universe with bulk viscous fluid in Bianchi V space-time}, Fortschr. Phys. \textbf{69}, 2100007 (2021).

\bibitem{sahni_statefinder_2003}
V.~Sahni, T.~D.~Saini, A.~A.~Starobinsky, U.~Alam, \textit{Statefinder---a New Geometrical Diagnostic of Dark Energy}, JETP Lett.\ \textbf{77}, 201--206 (2003).

\bibitem{alam_exploring_2003}
U.~Alam, V.~Sahni, T.~D.~Saini, A.~A.~Starobinsky, \textit{Exploring the Expanding Universe and Dark Energy Using the Statefinder Diagnostic}, Mon.\ Not.\ R.\ Astron.\ Soc.\ \textbf{344}, 1057--1074 (2003).

\bibitem{krssak_ft_2019}
M.~Krssak et al., \textit{Teleparallel theories of gravity: illuminating a fully invariant approach}, Classical and Quantum Gravity \textbf{36}, 183001 (2019).

\bibitem{mandal_temporal_2020}
S.~Mandal, P.~K.~Sahoo, \textit{On The Temporal Evolution of Particle Production in $f(T)$ Gravity}, Mod. Phys. Lett. A \textbf{35}, 2050328 (2020).

\bibitem{husain_logamediate_2026}
A.~Husain, S.~Khan, R.~Agrawal, S.~Shekh, \textit{Logamediate expansion and dark-energy-driven acceleration in $f\left(T\right)$ teleparallel gravity}, Indian J Phys (2026).

\bibitem{duchaniya_dynamical_2022}
L.~K.~Duchaniya, S.~V.~Lohakare, B.~Mishra, S.~K.~Tripathy, \textit{Dynamical stability analysis of accelerating f(T) gravity models}, Eur. Phys. J. C \textbf{82}, 448 (2022).

\bibitem{chaudhary_constraints_2023}
H.~Chaudhary et al., \textit{Constraints on the parameters of modified Chaplygin-Jacobi and modified Chaplygin-Abel gases in $f(T)$ gravity}, preprint, doi:10.21203/rs.3.rs-3656086/v1 (2023).

\bibitem{sharif_dark_2014}
M.~Sharif, S.~Azeem, \textit{Dark Energy Models and Cosmic Acceleration with Anisotropic Universe in $f(T)$ Gravity}, Commun. Theor. Phys. \textbf{61}, 482--490 (2014).

\bibitem{zubair_thermodynamic_2015}
M.~Zubair, S.~Waheed, \textit{Thermodynamic study in modified  $f(T)$ gravity with cosmological constant regime}, Astrophys Space Sci \textbf{360}, 68 (2015).

\bibitem{setare_cosmological_2012}
M.~R.~Setare, N.~Mohammadipour, \textit{Cosmological viability conditions for $f(T)$ dark energy models}, J. Cosmol. Astropart. Phys. \textbf{2012}, 030--030 (2012).

\bibitem{jimenez_constraints_2002}
R.~Jimenez, A.~Loeb, \textit{Constraining Cosmological Parameters Based on Relative Galaxy Ages}, Astrophys.\ J.\ \textbf{573}, 37--42 (2002).

\bibitem{moresco_raising_2016}
M.~Moresco et al., \textit{Raising the bar: new constraints on the Hubble parameter with cosmic chronometers at $z\sim2$}, J.\ Cosmol.\ Astropart.\ Phys.\ \textbf{2016}, 014 (2016).

\bibitem{nunes_new_2017}
R.~C.~Nunes, S.~Pan, E.~N.~Saridakis, E.~M.~C.~Abreu, \textit{New observational constraints on $f(R)$ gravity from cosmic chronometers}, J. Cosmol. Astropart. Phys. \textbf{2017}, 005--005 (2017).

\bibitem{sudharani_probing_2024}
L.~Sudharani, N.~S.~Kavya, D.~M.~Naik, V.~Venkatesha, \textit{Probing accelerating cosmos via reconstructed Hubble parameter and its influence on $f(T)$ gravity models}, Nuclear Physics B \textbf{998}, 116410 (2024).

\bibitem{brout_pantheon_2022}
D.~Brout et al., \textit{The Pantheon$+$ Analysis: Cosmological Constraints}, Astrophys.\ J.\ \textbf{938}, 110 (2022).

\bibitem{scolnic_pantheon_2022}
D.~Scolnic et al., \textit{The Pantheon$+$ Analysis: The Full Data Set and Light-Curve Release}, Astrophys.\ J.\ \textbf{938}, 113 (2022).

\bibitem{beutler_6df_2011}
F.~Beutler et al., \textit{The 6dF Galaxy Survey: baryon acoustic oscillations and the local Hubble constant}, Mon.\ Not.\ R.\ Astron.\ Soc.\ \textbf{416}, 3017--3032 (2011).

\bibitem{alam_boss_2017}
S.~Alam et al., \textit{The clustering of galaxies in the completed SDSS-III BOSS: cosmological analysis of the DR12 galaxy sample}, Mon.\ Not.\ R.\ Astron.\ Soc.\ \textbf{470}, 2617--2652 (2017).

\end{thebibliography}
\end{document}